\documentclass[12pt]{scrartcl} 
\usepackage{a4}
\usepackage{empheq}
\usepackage{amsfonts}
\usepackage{mathrsfs}
\usepackage{bbm}
\usepackage{booktabs}
\usepackage[all,cmtip]{xy}
\usepackage[a4paper,colorlinks,citecolor=blue,linkcolor=blue,urlcolor=blue,pdfpagemode=None]{hyperref} 
\usepackage[square,numbers,sort&compress]{natbib}

\vfuzz \hfuzz
\newcommand{\I}{\text{i}}
\newcommand{\C}{\mathbbm{C}}

\newcommand{\N}{\mathbbm{N}}

\newcommand{\Z}{\mathbbm{Z}}
\newcommand{\one}{\mathbbm{1}}
\newcommand{\Ainf}{A_\infty}

\allowdisplaybreaks

\deffootnote[1em]{1em}{1em}{\textsuperscript{\thefootnotemark}}

\begin{document}

\title{Triangle-generation in\\ topological D-brane categories}
\author{Nils Carqueville
\\[0.5cm]
  {\normalsize\slshape King's College London, Department of Mathematics,}\\[-0.1cm]
  {\normalsize\slshape Strand, London WC2R\,2LS, UK}\\[-0.1cm]
  \normalsize{\tt \href{mailto:nils.carqueville@kcl.ac.uk}{nils.carqueville@kcl.ac.uk}}}
\date{}
\maketitle

\begin{abstract}
Tachyon condensation in topological Landau-Ginzburg models can generally be studied using methods of commutative algebra and properties of triangulated categories. The efficiency of this approach is demonstrated by explicitly proving that every D-brane system in all minimal models of type ADE can be generated from only one or two fundamental branes. 
\end{abstract}

\subsection*{Introduction}

Formulating the study of D-branes in B-twisted topological Landau-Ginzburg models in terms of matrix factorisations has proved to be very useful. The interplay of an elegant and abstract mathematical setting on the one hand and its explicit, simple manifestations on the other hand has been applied in numerous instances to both prove general theorems and carry out concrete computations. Prominent examples for this are the large-volume limit/stringy regime dualities of~\cite{o0302,o0503,qv,hhp} which are open string counterparts and generalisations of the construction of~\cite{w9301}, or the many links of topological D-branes described by conformal field theory to their Landau-Ginzburg analogues, see e.\,g.~\cite{bg0503,bg0506,kl0306,h0401,err0508,ks0610175}. 

A field in which this complementary interplay has not been exploited to its full potential is in analysing topological tachyon condensation. The fact that the dynamics of forming D-brane composites is an inherently off-shell process suggests that this type of tachyon condensation should best be studied within the framework of open string field theory. This is emphasised by the significance of Sen's conjectures~\cite{s9911116,s9805170} and their partial proofs so far following~\cite{s0511286}, as well as by the conceptual clarity of the approach of~\cite{l0102122,l0102183} in terms of generalised complexes in differential graded categories (which are $A_\infty$-quasi-equivalent to any open string field theory~\cite{l0310337}). 

On the other hand, it was argued in~\cite{d0011,al0104} that important aspects of tachyon condensation in open topological string theory can already be studied at the on-shell level. The latter is believed to be modelled by the structure of triangulated categories. In this setting, a particular D-brane composite is obtained by deforming the superposition of two branes by a potentially tachyonic string, and this type of tachyon condensation corresponds exactly to the cone construction in the triangulated category. Detailed studies of such processes in topological Landau-Ginzburg models were only undertaken in special cases by making use of additional structure that is not generally available. Examples include models with superpotentials in only one variable, where the special properties of the underlying polynomial ring can be utilised~\cite{hll0405}, or cases where further (geometric) dualities are available and under control~\cite{bhlw0408,gjlw0512,lpp2002}. 

\vspace{0.2cm}

In the present note a general and practical method of computing ``double-cones'' to analyse topological tachyon condensation is proposed. It builds on the triangulated structure of matrix factorisations as well as basic notions and results in commutative algebra that are readily implemented in computer algebra systems. Part of this approach is an algorithmic computation of BRST-cohomology in Landau-Ginzburg models which is also of independent interest, e.\,g.~in the study of deformation theory. 

After a general discussion of this simple yet effective method in section~\ref{Preliminaries}, its power is illustrated in section~\ref{ADE} by proving that, in topological Landau-Ginzburg models with potentials of type ADE, any D-brane can be produced from tachyon condensations of D-branes of only one single fundamental type, except for the case of type D$_{\ell}$ with $\ell$ even where two fundamental branes are needed. In other words, the associated categories of matrix factorisations are each triangle-generated by one or two indecomposable objects. This is consistent with the results of~\cite{kst0511} whose authors use the special properties of the associated ADE Auslander-Reiten quivers. While this approach is very elegant, it is not as generally available as the approach discussed here, which can also easily provide further details such as proving that models of type D$_{2n+1}$ are generated by a single indecomposable object. 

The results for models of type A allow for immediate comments on two related issues. The relation to the corresponding generation result of~\cite{h0401} for type A models is given by reformulating it in terms of the precise language of triangulated categories (which however do not seem to capture the continuous nature of the ``homotopy flows'' of~\cite{h0401}) and their interpretation. Furthermore, first steps of discussing bulk-induced renormalisation group flow in terms of matrix factorisations are taken. 

As the complete discussion of E-type models involves quite a number of explicit matrices, the details of this analysis are relegated to an appendix. The sheer number of rather large matrices is aesthetically unpleasing, but at the same time it demonstrates the favourable explicit control over the details of the category of matrix factorisations.

\section{General strategy}\label{Preliminaries}

\subsection*{Preliminaries}

D-branes in B-twisted topological Landau-Ginzburg models with potential $W$ are described by the category of matrix factorisations $\text{MF}(W)$~\cite{kl0210,bhls0305,l0312}. Its objects, which model the D-branes, are polynomial square matrices of block form $Q=(\begin{smallmatrix}0&g\\ f&0\end{smallmatrix})$ such that $fg=gf=W\one$. This property implies that for any two matrix factorisations $Q$ and $Q'$, the map defined as the linear extension of the map $\phi\mapsto Q'\phi-(-1)^{|\phi|}\phi Q$ for homogeneous $\phi$ is a differential $D_{QQ'}$ on the $\Z_2$-graded space of polynomial matrices of fixed size, where block-diagonal matrices have degree 0 and off-block-diagonal matrices have degree 1. The morphisms between $Q$ and $Q'$, which describe the bosonic topological strings between the corresponding branes, are defined to be the $0$-th cohomology $H^0(Q,Q')$ of $D_{QQ'}$. Fermionic states are not left behind as $H^1(Q,Q')$ is isomorphic to $H^0(Q,\bar Q')$ where $\bar Q'=-(\begin{smallmatrix}0&g'\\ f'&0\end{smallmatrix})$ is the anti-brane to $Q'$. More precisely, an element $(\begin{smallmatrix}0&f_1\\ f_0&0\end{smallmatrix})$ is in $H^1(Q,Q')$ iff $(\begin{smallmatrix}f_0&0\\ 0&f_1\end{smallmatrix})$ is in $H^0(Q,\bar Q')$. 

The category $\text{MF}(W)$ has much more structure that admits a physical interpretation; it is even believed that matrix factorisations have the full structure of open topological string theories. While in general this is still a conjecture awaiting proof (by constructing a suitable minimal, unital and cyclic $A_\infty$-structure on the category of matrix factorisations, see~\cite{hll0402,c0412149}), one can straightforwardly prove that $\text{MF}(W)$ is a triangulated category~\cite{Neeman,o0302,GM}. In particular, it is endowed with a shift functor, which corresponds to the transition to the anti-brane of a given brane, and with distinguished triangles. These are sequences of the form
$$
\xymatrix{Q \ar[r]^{\varphi} & Q' \ar[r]^{} & \text{C}(\varphi) \ar[r]^{} & \bar Q}
$$
and are believed to describe the result of topological tachyon condensation as the mapping cone $\text{C}(\varphi)$ defined as
$$
\text{C}(\begin{smallmatrix} \varphi_0 & 0 \\ 0 & \varphi_1 \end{smallmatrix}) \equiv \text{C}\big((\begin{smallmatrix} \varphi_0 & 0 \\ 0 & \varphi_1 \end{smallmatrix}):Q\longrightarrow Q'\big) = \begin{pmatrix}0&0&-f&0\\ 0&0&\varphi_0&g'\\ -g&0&0&0\\ \varphi_1&f'&0&0\end{pmatrix} \, .
$$
This definition is mathematically natural in the sense that it is formally the same as in the case of $\Z$-graded complexes, see e.\,g.~\cite{GM}, and the resulting matrix factorisation $\text{C}(\begin{smallmatrix} \varphi_0 & 0 \\ 0 & \varphi_1 \end{smallmatrix})$ is again of the form $(\begin{smallmatrix} 0 & G \\ F & 0 \end{smallmatrix})$. An equivalent definition is to introduce the cone as $(\begin{smallmatrix} Q & 0 \\ \psi & Q' \end{smallmatrix})$ for fermionic strings $\psi$. 

The precise relation between cones $\text{C}(\varphi)$ and tachyon condensation signalled by the corresponding string is as follows. By one of the axioms of triangulated categories, every morphism $\varphi:Q\rightarrow Q'$ is part of a distinguished triangle $Q\rightarrow Q'\rightarrow\text{C}(\varphi)\rightarrow\bar Q$, and this is interpreted in such a way that the branes described by $Q$ and $\text{C}(\varphi)$ may bind to form $Q'$, or, equivalently, $\bar Q$ and $Q'$ may condense to form $\text{C}(\varphi)$. It is precisely this process that is captured in the notion of Grothendieck groups. These K-theoretic groups are believed to classify topological D-brane charges and can be defined for an arbitrary triangulated category~$\mathcal T$: The Grothendieck group $K_0(\mathcal T)$ is the free abelian group of isomorphism classes $[X]$ of objects $X$ in $\mathcal T$ modulo the relations $[X]-[Y]+[Z]=0$ for all distinguished triangles $X\rightarrow Y\rightarrow Z\rightarrow\bar X$ in $\mathcal T$.

\subsection*{Tachyon condensation and double-cones}

A physically interesting question to ask is whether a given D-brane system can be generated via tachyon condensation of more fundamental branes. With the assumed description in terms of triangulated categories this can be translated into the following well-posed mathematical question: Is there a set of ``fundamental'' objects $\{Q_i\}_{i\in I}$ in $\text{MF}(W)$ that triangle-generates a given triangulated subcategory, i.\,e.~is this subcategory contained in $\text{tria}(\{Q_i\}_{i\in I})$, the closure of the iterated operations of taking cones and shifts of the fundamental objects $Q_i$? This simple reformulation offers a number of ways to check whether a given system of D-branes can be the result of topological tachyon condensation of some other branes. The generally applicable method that will be used in this note utilises the basic fact that a morphism in a triangulated category is an isomorphism iff its cone is isomorphic to zero~\cite{Neeman}. To see this, let $\varphi:Q\rightarrow Q'$ be an isomorphism. Then the triangle $Q\stackrel{\varphi}{\rightarrow}Q'\rightarrow 0\rightarrow \bar Q$ is isomorphic to the triangle $Q'\stackrel{1}{\rightarrow}Q'\rightarrow 0\rightarrow \bar Q'$ which is distinguished by definition. Hence $\text{C}(\varphi)\cong 0$ as the third object in a distinguished triangle is unique up to isomorphisms. Conversely, if $Q\stackrel{\varphi}{\rightarrow}Q'\rightarrow 0\rightarrow \bar Q$ is a distinguished triangle, then
$$
\xymatrix{Q \ar[r]^{\varphi} \ar[d]_{\varphi} & Q' \ar[r]^{} \ar[d]_{1} & 0 \ar[r]^{} \ar[d]_{0} & \bar Q \ar[d]_{\bar\varphi} \\
Q' \ar[r]^{1} & Q' \ar[r]^{} & 0 \ar[r]^{} & \bar Q'}
$$
is a morphism of distinguished triangles. But as the vertical maps $1=\one_{Q'}$ and~$0$ are isomorphisms, $\varphi$ also must be an isomorphism. 

Using this result one can now check whether a string described by $\varphi\in H^0(Q,Q')$ leads to a condensation to another brane $Q''$ by considering all morphisms $\psi$ between $Q''$ and $\text{C}(\varphi)$. If for one such $\psi$ its cone has zero endomorphism space, then $\psi$ is an isomorphism and hence $Q''$ can be generated from $\bar Q$ and $Q'$. So once the computation of morphism spaces is under control, this simple idea provides a practical tool to study tachyon condensation in general topological D-brane categories. For the case of Landau-Ginzburg models, a straightforward algorithm to compute arbitrary open string BRST-cohomologies will be described in the next subsection. 

The method just presented works under very general assumptions. In particular, without knowing the complete and intricate structure of $\text{MF}(W)$, one has a tool to decide whether certain ``interesting'' branes can be generated from certain ``fundamental'' branes. A special case of this is when ``interesting'' means ``all'' and $\text{MF}(W)$ is known to be semi-simple or completely decomposable such that the ``fundamental'' branes can be taken to correspond to a subset of indecomposable objects. This situation is realised in the case of Landau-Ginzburg models with potentials of singularity type ADE which correspond to minimal $N=2$ superconformal field theories, and which will be studied in detail in section~\ref{ADE}. 

On the other hand, even if the model at hand is more complicated and extensive knowledge of decomposability is not available, the method proposed here is still useful by restricting to a particular ``interesting'' subcategory. For instance, many Landau-Ginzburg potentials that allow for non-trivial tensor product or linear matrix factorisations are important examples of cases that are not known to have a similarly simple structure as the minimal models, and hence similarly complete analyses are not possible at the moment. Nevertheless, the method of computing cohomologies for double-cones can still be applied to study for example tachyon condensation of permutation branes, even though a complete classification of all matrix factorisations of the associated potentials is not known. In fact, such applications may help to gain new insight into the inner structure of more complicated D-brane categories.

\subsection*{Algorithmic computation of cohomology}

To utilise the method described above efficiently, it is convenient (and in practice mostly indispensable) to compute the morphism spaces $H^0(Q,Q')$ algorithmically and in an automatised fashion. With the help of computer algebra systems like Singular~\cite{Singular} and basic notions and results of commutative algebra this is straightforwardly accomplished as will now be explained.\footnote{The approach taken here is entirely different from and more simple-minded than the work of~\cite{ag2001} as discussed in~\cite{a0703279}. According to P.\,S.~Aspinwall, the code of~\cite{ag2001} may also be used to compute explicit cohomology representatives, but only after modifying it suitably.} The reader is referred to~\cite{SingularBook,eisenbudcommalg,MFcohom} for definitions and background on elementary notions in commutative algebra used below and to~\cite{MFcohom} for a detailed account of the implementation together with additional examples and applications. 

Let $Q=(\begin{smallmatrix}0&g\\ f&0\end{smallmatrix})$ and $Q'=(\begin{smallmatrix}0&g'\\ f'&0\end{smallmatrix})$ be two $2r\times 2r$ matrix factorisations of $W=W(x_1,\ldots,x_N)\equiv W(X)$. The differential $D_{QQ'}$ decomposes into the sum of $D^-_{QQ'}=D_{QQ'}\pi_-$ and $D^+_{QQ'}=D_{QQ'}\pi_+$, where $\pi_-$ and $\pi_+$ project on the subspaces of $\Z_2$-degree $0$ and $1$, respectively. As the differentials $D^\pm_{QQ'}$ are linear maps on the space of polynomial $2r\times 2r$ matrices, they can be represented as matrices
$$
D^- = \begin{pmatrix}A&B\\C&D\end{pmatrix} \quad \text{and} \quad D^+ = \begin{pmatrix}-D&+B\\+C&-A\end{pmatrix} \, ,
$$
respectively, in $\text{Mat}(\C[X],2r^2)$,\footnote{Actually, $D^\pm_{QQ'}$ are canonically isomorphic to elements in $\text{Mat}(\C[X],4r^2)$, but because of their homogeneous $\Z_2$-degree +1 half of the entries of the column vectors in their image are always zero, so that they can be viewed as elements in $\text{Mat}(\C[X],2r^2)$, see also~\cite{MFcohom}.} where the entries of $A,B,C,D\in\text{Mat}(\C[X],r^2)$ are given by
\begin{subequations}
\begin{align*}
A_{ij} & = - \sum_{k=0}^{r-1} \sum_{l,m=1}^r \delta_{i,kr+l}\, \delta_{j,kr+m}\, g_{ml} \, , && B_{ij} = \sum_{l,k=0}^{r-1} \sum_{m=1}^r \delta_{i,lr+m}\, \delta_{j,kr+m}\, g'_{l+1,k+1} \, , \\
C_{ij} & = \sum_{l,k=0}^{r-1} \sum_{m=1}^r \delta_{i,lr+m}\, \delta_{j,kr+m}\, f'_{l+1,k+1} \, , && D_{ij}  = - \sum_{k=0}^{r-1} \sum_{l,m=1}^r \delta_{i,kr+l}\, \delta_{j,kr+m}\, f_{ml} \, .
\end{align*}
\end{subequations}
The cohomology $H^0(Q,Q')$ is isomorphic to $\text{Ker}D^-/\text{Im}D^+$. To compute this, first note that $\text{Im}D^+_{QQ'}$ is simply isomorphic to the $\C[X]$-module generated by the columns of the matrix $D^+$, where the isomorphism is given by the translation from matrices to column vectors on which the matrices $D^\pm$ act. 

Secondly, the computation of $\text{Ker}D^-_{QQ'}\cong\text{Ker}D^-$ is a syzygy problem which can be solved by standard methods. Indeed, let $D^-$ be given by its column vectors $d_i$ and let $\{e_i\}$ denote the canonical basis of $\C^{2r^2}$. Then $\varphi=\sum_{i=1}^{2r^2}\varphi_i e_i$, with $\varphi_i\in\C[X]$, is an element of $\text{Ker}D^-$ iff $D^-\varphi=\sum_{i=1}^{2r^2}\varphi_i d_i=0$ which means that $(\varphi_1,\ldots,\varphi_{2r^2})\in\C[X]^{2r^2}$ is a syzygy for $(d_1,\ldots,d_{2r^2})\equiv D^-$. The computation of syzygies is implemented in Singular using the Gr\"{o}bner basis algorithm~\cite{buchberger,SingularBook}, so the computation of a finite set of generators of the (infinite-dimensional) $\C[X]$-module $\text{Ker}D^-$ can be automatised. 

Thirdly, one has to find those elements of $\text{Ker}D^-$ which are not in $\text{Im}D^+$. This is a ``module membership problem''  which can also be solved algorithmically using Singular. To do this, a version of Buchberger's algorithm~\cite{buchberger} is used to compute a (reduced) normal form on $\C[X]^{2r^2}$, and subsequently the following basic result in commutative algebra is applied: if $\text{NF}(\,\cdot\,|\,\cdot\,)$ is a weak normal form and $B$ is a standard basis of a submodule $M\subset\C[X]^{2r^2}$, then an arbitrary element $\varphi$ in $\C[X]^{2r^2}$ is in $M$ iff $\text{NF}(\varphi|B)=0$. 

Hence one may begin by checking whether the first generator $G_1$ of $\text{Ker}D^-$ (obtained by solving the syzygy problem) is in $\text{Im}D^+$. If this is the case, the submodule generated by $G_1$ is in $\text{Im}D^+$, and $G_1$ gives no contribution to the cohomology. But if Singular concludes that $G_1$ is not in $\text{Im}D^+$, it can be taken as a first explicit representative~$k_{11}$ of a basis element of the cohomology, $k_{11}=G_1$. In this case, one proceeds by checking whether elements of the form $x_i G_1$, $x_ix_j G_1$~etc.~are in $\text{Im}D^+$. After a finite number of steps (assuming that BRST-cohomology is finite-dimensional) one finds the finite list of basis vectors $k_{1j}$ (over $\C$) in the module $\C[X]G_1$ whose equivalence classes $\overline{k}_{1j}$ are elements of the cohomology $\text{Ker}D^-/\text{Im}D^+$, and the same procedure is repeated for the other submodules $\C[X]G_i$, where $G_i$ are the remaining generators of $\text{Ker}D^-$, to determine all representatives $k_{ij}$.

Finally, one needs to check whether all $\overline{k}_{ij}$ previously obtained are linearly independent. This is not guaranteed even though the $k_{ij}$ are linearly independent. To perform the check, one may first compare $\sharp\{k_{ij}\}$ with the actual dimension of $\text{Ker}D^-/\text{Im}D^+$, which is determined in Singular by computing a presentation matrix for the quotient module $\text{Ker}D^-/\text{Im}D^+$, i.\,e.~a matrix whose cokernel is isomorphic to the module. For many important pieces of information such as the dimension, the knowledge of this presentation is completely sufficient. But in order to find explicit representatives of cohomology basis elements, a slightly more involved treatment like the one described here is necessary. 

If $\sharp\{k_{ij}\}$ is equal to the dimension of $\text{Ker}D^-/\text{Im}D^+$, $\{\overline{k}_{ij}\}$ can be taken as a basis of $\text{Ker}D^-/\text{Im}D^+$. But if $\sharp\{k_{ij}\}>\text{dim}(\text{Ker}D^-/\text{Im}D^+)$, one has to find a maximal linearly independent subset of $\{\overline{k}_{ij}\}$ in the quotient module. With the help of the properties of normal forms, this task can be reduced to the corresponding standard problem of determining maximal linearly independent subsets before quotienting, and the computation is completed.

\section{Tachyon condensation in models of type ADE}\label{ADE}

The ideas of the previous section can particularly easily be applied to those Landau-Ginzburg models whose infra-red fixed points under RG flow are identified with minimal $N=2$ superconformal field theories~\cite{m1989,vw1989,lvw1989}. Their superpotentials $W$ are equal (up to adding squares of new variables) to one of the ADE-type polynomials listed in table~\ref{ADEGrothendieck}. 
\begin{table}[t]
\begin{center}
\begin{tabular}{lll}
\toprule[2pt]
& ADE polynomial $W$ & $K_0(\text{MF}(W))$\tabularnewline
\midrule[1.5pt]
A$_n$: & $x^{n+1}$ & $\Z_{n+1}$\\
\addlinespace
D$_n$: & $x^2y+y^{n-1}+z^2$ & $\Z_2\oplus\Z_2$\phantom{$\Z_4$}\!\!\!\!\!\! for $n$ even\\
&& $\Z_4$\phantom{$\Z_2\oplus\Z_2$}\!\!\!\!\!\! for $n$ odd\\
\addlinespace
E$_6$: & $x^3+y^4+z^2$ & $\Z_3$\\
E$_7$: & $x^3+xy^3+z^2$ & $\Z_2$\\
E$_8$: & $x^3+y^5+z^2$ & $	\{0\}$\tabularnewline
\bottomrule[2pt]
\end{tabular}%
\caption{ADE polynomials and their associated Grothendieck groups.\label{ADEGrothendieck}}%
\end{center}%
\end{table}%
For these models the category $\text{MF}(W)$ is completely decomposable and the indecomposable objects are explicitly known. In this section it is shown that in each case $\text{MF}(W)$ is triangle-generated by only one of these fundamental objects, unless $W$ is of type D$_\ell$ with $\ell$ even where two such objects are needed. This is interpreted in such a way that any D-brane system in these models can be viewed as a tachyon condensate of (copies of) only one or two fundamental branes. 

To prove this for types D and E, the method of computing double-cones of the previous section is applied, while in the easier case of type A an even simpler method is available. For the latter case the discussion here essentially reproduces results of~\cite{h0401,hll0405}, though in these references the triangulated structure is not made use of. 

\vspace{0.2cm}

Table~\ref{ADEGrothendieck} also collects the Grothendieck groups $K_0(\text{MF}(W))$ of the categories of matrix factorisations of polynomials $W$ of type ADE (which are computed for example in~\cite{yoshino,schreyer1987}). Knowledge of these explicit expressions will be helpful for some of the arguments in this section. 

It should be noted that apart from the ADE polynomials listed in table~\ref{ADEGrothendieck}, there are also versions of them where the square $z^2$ is discarded (respectively added in the A-type case). The corresponding models are those with an opposite GSO projection, and their treatment is the same apart from straightforward adaptations. This is why the following analysis will only deal with the former cases.

\subsection{ADE-type A}

It is shown e.\,g.~in~\cite{o0302} (see also~\cite{hll0405}) that any object in $\text{MF}(x^n)$ is isomorphic to a finite direct sum of the indecomposable objects $Q_a=(\begin{smallmatrix} 0 & x^{n-a} \\ x^a & 0 \end{smallmatrix})$ with $a\in\{1,\ldots,n-1\}$. The bosonic and fermionic open string spaces $H^0(Q_a,Q_b)$ and $H^1(Q_a,Q_b)$ between D-branes $Q_a$ and $Q_b$ are given by the cohomologies
\begin{align*}
& \C \left\{ \begin{pmatrix} x^{i+a-b} & 0 \\ 0 & x^i \end{pmatrix} \;\Big|\; \text{max}(b-a,0)\leq i \leq \text{min}(b-1,n-a-1) \right\} \, , \\
& \C \left\{ \begin{pmatrix} 0 & x^{n-a-b+i} \\ -x^i & 0 \end{pmatrix} \;\Big|\; \text{max}(a+b-n,0)\leq i \leq \text{min}(b-1,a-1) \right\} \, ,
\end{align*}
respectively. 

The task now is to start with one single D-brane and its anti-brane, compute the mapping cones for all their morphisms and find similarity transformations (which are special instances of isomorphisms in $\text{MF}(x^n)$) such that these cones are related to direct sums of indecomposable matrix factorisations. It turns out that some cones are isomorphic in this way to a direct sum of a trivial matrix factorisation of the form $(\begin{smallmatrix} 0 & 1 \\ x^n & 0 \end{smallmatrix})$ or $(\begin{smallmatrix} 0 & x^n \\ 1 & 0 \end{smallmatrix})$ and another indecomposable object. All indecomposable D-branes can be produced this way. 

Whenever the Landau-Ginzburg potential $W$ is a polynomial of only one single variable, one can find a unimodular polynomial matrix $U=(\begin{smallmatrix} S & 0 \\ 0 & T \end{smallmatrix})$ such that $U\text{C}(\begin{smallmatrix} \varphi_0 & 0 \\ 0 & \varphi_1 \end{smallmatrix})U^{-1}$ is of the form $(\begin{smallmatrix} 0 & q_1 \\ q_0 & 0 \end{smallmatrix})$ with $q_0=\text{diag}(p_1,p_2)$ and $q_1=\text{diag}(W/p_1,W/p_2)$ where $p_1$ and $p_2$ are polynomials that have certain divisibility properties. The unique diagonal matrix is called the associated Smith normal form which exists in any principal integral domain and there is a straightforward algorithm to construct it (see e.\,g.~\cite{smithnormalform}). 

With the procedure just outlined it is possible to compute all cones and relate them to direct sums of indecomposable objects for the potential $x^n$ for any fixed~$n$: by computing the Smith form one determines the monomials $p_1=x^i$ and $p_2=x^j$ for some $i,j\in\N$, and this means that the cone under consideration is isomorphic to the sum of the objects $Q_i$ and $Q_j$.  In order to get an idea of the structure, the results for the special case of $n=5$ are listed in table~\ref{x5cones}, where the box in the $i$-th row and $j$-th column contains the cones for all morphisms between $Q_i$ and $Q_j$ as well as their decomposition into direct sums of fundamental objects. $Q_0$ and $Q_5$ are isomorphic to zero objects in $\text{MF}(x^n)$ and can thus be deleted in any decomposition where they appear. 
\begin{table}[t]
\begin{center}
{\footnotesize
\begin{tabular}{@{}*{5}{l}@{}}
\toprule[2pt]
& $j=1$ & $j=2$ & $j=3$ & $j=4$\tabularnewline
\midrule[1.5pt]
$i=1$ & $\text{C}(\begin{smallmatrix} 1 & 0 \\ 0 & 1 \end{smallmatrix})\cong Q_0\oplus Q_5$ & $\text{C}(\begin{smallmatrix} 1 & 0 \\ 0 & x \end{smallmatrix})\cong Q_1\oplus Q_5$ & $\text{C}(\begin{smallmatrix} 1 & 0 \\ 0 & x^2 \end{smallmatrix})\cong Q_2\oplus Q_5$ & $\text{C}(\begin{smallmatrix} 1 & 0 \\ 0 &x^3 \end{smallmatrix})\cong Q_3\oplus Q_5$ \\
\addlinespace
$i=2$ & $\text{C}(\begin{smallmatrix} x & 0 \\ 0 & 1 \end{smallmatrix})\cong Q_0\oplus Q_4$ & $\text{C}(\begin{smallmatrix} 1 & 0 \\ 0 & 1 \end{smallmatrix})\cong Q_0\oplus Q_5$ & $\text{C}(\begin{smallmatrix} 1 & 0 \\ 0 & x \end{smallmatrix})\cong Q_1\oplus Q_5$ & $\text{C}(\begin{smallmatrix} 1 & 0 \\ 0 &x^2 \end{smallmatrix})\cong Q_2\oplus Q_5$ \\
& & $\text{C}(\begin{smallmatrix} x & 0 \\ 0 & x \end{smallmatrix})\cong Q_1\oplus Q_4$ & $\text{C}(\begin{smallmatrix} x & 0 \\ 0 & x^2 \end{smallmatrix})\cong Q_2\oplus Q_4$ & \\
\addlinespace
$i=3$ & $\text{C}(\begin{smallmatrix} x^2 & 0 \\ 0 & 1 \end{smallmatrix})\cong Q_0\oplus Q_3$ & $\text{C}(\begin{smallmatrix} x & 0 \\ 0 & 1 \end{smallmatrix})\cong Q_0\oplus Q_4$ & $\text{C}(\begin{smallmatrix} 1 & 0 \\ 0 & 1 \end{smallmatrix})\cong Q_0\oplus Q_5$ & $\text{C}(\begin{smallmatrix} 1 & 0 \\ 0 &x \end{smallmatrix})\cong Q_1\oplus Q_5$ \\
& & $\text{C}(\begin{smallmatrix} x^2 & 0 \\ 0 & x \end{smallmatrix})\cong Q_1\oplus Q_3$ & $\text{C}(\begin{smallmatrix} x & 0 \\ 0 & x \end{smallmatrix})\cong Q_1\oplus Q_4$ & 
\\
\addlinespace
$i=4$ & $\text{C}(\begin{smallmatrix} x^3 & 0 \\ 0 & 1 \end{smallmatrix})\cong Q_0\oplus Q_2$ & $\text{C}(\begin{smallmatrix} x^2 & 0 \\ 0 & 1 \end{smallmatrix})\cong Q_0\oplus Q_3$ & $\text{C}(\begin{smallmatrix} x & 0 \\ 0 & 1 \end{smallmatrix})\cong Q_0\oplus Q_4$ & $\text{C}(\begin{smallmatrix} 1 & 0 \\ 0 &1 \end{smallmatrix})\cong Q_0\oplus Q_5$ \\
\bottomrule[2pt]
\end{tabular}%
}%
\caption{Cones of $Q_i\rightarrow Q_j$ in $\text{MF}(x^5)$.\label{x5cones}}%
\end{center}%
\end{table}%

The table shows that one can for example start with the D-brane $Q_1$ and its anti-brane $Q_4$ and then compute the tachyon condensation induced by the only string state between them,~i.e.~the cone $\text{C}(\begin{smallmatrix} 1 & 0 \\ 0 & x^3 \end{smallmatrix})$. This cone is isomorphic to the object  $Q_3$ (the similarity transformation is given by $S=(\begin{smallmatrix} 1 & x \\ 0 & 1 \end{smallmatrix})$ and $T=(\begin{smallmatrix} 0 & 1 \\ 1 & x \end{smallmatrix})$ in the above notation), and by adding its anti-brane $Q_2$, one already has found all indecomposable objects in this case. Note that the table also shows that one may start with any other indecomposable D-brane and its anti-brane.

For arbitrary $n$, one can always take the above as the first step in the construction of all indecomposable objects and then proceed inductively: 
\begin{enumerate}
\item Assume that via tachyon condensation of a system of D-branes of a single type one can produce the indecomposable objects 
$$
Q_1, Q_2, \ldots, Q_{2k-2}, Q_{2k-1}; Q_{n-2k+1}, Q_{n-2k+2}, \ldots, Q_{n-2}, Q_{n-1},
$$
where $k\geq 2$ and $n\geq 4k$ (so that there is at least one D-brane left to construct). Then $(\begin{smallmatrix} 1 & 0 \\ 0 & x^{n-2k} \end{smallmatrix})$ is an element of $H^0(Q_k,Q_{n-k})$ and one can show that $U_{2k} \text{C}(\begin{smallmatrix} 1 & 0 \\ 0 & x^{n-2k} \end{smallmatrix}) U_{2k}^{-1}$ is equal to
$$
\begin{pmatrix}0&0&x^{2k}&0\\ 0&0&0&1\\ x^{n-2k}&0&0&0\\ 0&x^n&0&0\end{pmatrix} \cong Q_{n-2k} \quad\text{with}\quad U_{2k}=\begin{pmatrix}1&x^k&0&0\\ 0&1&0&0\\ 0&0&0&1\\ 0&0&1&x^k\end{pmatrix} \, .
$$
Thus $Q_{2k}$ and its anti-brane $Q_{n-2k}$ can be added to the list of D-branes that can be constructed from the initial D-brane. 
\item Assume that via tachyon condensation of a system of D-branes of a single type one can produce the indecomposable objects 
$$
Q_1, Q_2, \ldots, Q_{2l-1}, Q_{2l}; Q_{n-2l}, Q_{n-2l+1}, \ldots, Q_{n-2}, Q_{n-1},
$$
where $l\geq 1$ and $n\geq 4l+2$ (so that there is at least one D-brane left to construct). Then $(\begin{smallmatrix} 1 & 0 \\ 0 & x^{n-2l-1} \end{smallmatrix})$ is an element of $H^0(Q_l,Q_{n-l-1})$ and one can show that $U_{2l+1} \text{C}(\begin{smallmatrix} 1 & 0 \\ 0 & x^{n-2l-1} \end{smallmatrix}) U_{2l+1}^{-1}$ is equal to
$$
\begin{pmatrix}0&0&x^{2l+1}&0\\ 0&0&0&1\\ x^{n-2l-1}&0&0&0\\ 0&x^n&0&0\end{pmatrix} \cong Q_{n-2l-1} \quad\text{with}\quad U_{2l+1}=\begin{pmatrix}1&x^l&0&0\\ 0&1&0&0\\ 0&0&0&1\\ 0&0&1&x^{l+1}\end{pmatrix} \, .
$$
Thus $Q_{2l+1}$ and its anti-brane $Q_{n-2l-1}$ can be added to the list of D-branes that can be constructed from the initial D-brane.
\end{enumerate}
By iterating the steps (i) and (ii) alternately one can construct all indecomposable objects (and hence all D-branes) from the D-brane $Q_1$ or $Q_{n-1}$. Mathematically, this means that $\text{MF}(x^n)$ is triangle-generated by both $Q_1$ and $Q_{n-1}$ as taking shifts of objects is part of the triangle-generation process, while physically one can say that any topological D-brane in this type of models can be viewed as a condensate of a system of D-branes of just one type. An anti-brane does not have to be included in the physical condensation process as the anti-branes $Q_{n-2l-1}$ and $Q_{n-2k}$ are isomorphic (via the same similarity matrices $U_{2l+1}$ and $U_{2k}$) to the cones of the bosonic morphisms $(\begin{smallmatrix}  x^{n-2l-1} & 0 \\ 0 & 1 \end{smallmatrix})\in H^0(Q_{n-l-1},Q_l)$ and $(\begin{smallmatrix} x^{n-2k} & 0 \\ 0 & 1 \end{smallmatrix})\in H^0(Q_{n-k},Q_k)$, respectively, and these are isomorphic to fermionic morphisms in $H^1(Q_{n-l-1},Q_{n-l})$ and $H^1(Q_{n-k},Q_{n-k})$. 

While in the earlier example of $n=5$ any brane can be generated from tachyon condensations of copies of any fundamental brane $Q_i$, one should note that in general the brane/anti-brane pair $Q_1,Q_{n-1}$ is special with respect to its generating property. For instance if $n=6$, the D-brane $Q_3$ cannot generate the other fundamental branes individually but only non-trivial direct sums of them. By charge conservation, both cases are of course consistent with the general fact~\cite{h0401} that the charges of all branes in models of type A$_{n-1}$ take values in $K_0(\text{MF}(x^n))=\Z_n$. It follows from the above construction that the branes $Q_i$ can be assigned the charge $i\in\Z_n$. In the language of mapping cones and their interpretation in terms of tachyon condensation, the result for possible brane charges can also be obtained from the following observations (which together with the subsequent remarks on RG flow slightly digress from the main topic of the present paper).

\vspace{0.2cm}

\noindent\textbf{Relation to homotopy flows. } In~\cite{h0401} Hori essentially argues for a similar generating result. He finds ``homotopy relations'' of the form
$$
Q_{\ell_1} \oplus Q_{\ell_2} \simeq 
\begin{cases}
Q_{\ell_1+\ell_2} & \text{for } \ell_1+\ell_2\leq n, \\
Q_{\ell_1+\ell_2-n} & \text{for } \ell_1+\ell_2 > n,
\end{cases}
$$
which show that any fundamental D-brane $Q_\ell$ can be obtained by successive ``homotopy flows'' of $Q_1$. (Note that in~\cite{h0401} the notation $\mathscr{B}_L=Q_{L+1}$ is used.) For example, the relation $Q_{\ell_1} \oplus Q_{\ell_2} \simeq Q_{\ell_1+\ell_2}$ means that the left-hand side is equal to $(\begin{smallmatrix}  0 & g_0(x) \\ f_0(x) & 0 \end{smallmatrix})$ with $f_t(x)=R_t(\begin{smallmatrix} 1 & 0 \\ 0 & x^{\ell_2} \end{smallmatrix})R_t^{-1}(\begin{smallmatrix} x^{\ell_1-1} & 0 \\ 0 & 1 \end{smallmatrix})$, $g_t(x)=(\begin{smallmatrix} x^{n-\ell_1-\ell_2} & 0 \\ 0 & x^{n-\ell_2} \end{smallmatrix})R_t(\begin{smallmatrix} x^{\ell_2} & 0 \\ 0 & 1 \end{smallmatrix})R_t^{-1}$ and $R_t=(\begin{smallmatrix} \cos t & -\sin t \\ -\sin t & \cos t \end{smallmatrix})$, while the right-hand side is isomorphic to $(\begin{smallmatrix}  0 & g_{\pi/2}(x) \\ f_{\pi/2}(x) & 0 \end{smallmatrix})$. On the other hand, the precise physical interpretation of these continuous and periodic flows does not seem entirely clear, so it would be welcome to understand at least the relation of the endpoints of such flows in terms of mapping cones. Such a description is indeed available as can be shown by computing Smith forms as above: In the case $\ell_1+\ell_2\leq n$ the cone of the element $(\begin{smallmatrix}-1&0\\ 0&x^{n-\ell_1-\ell_2}\end{smallmatrix})\in H^0(Q_{\ell_1},\bar{Q}_{\ell_2})$ can be transformed into $Q_{n-\ell_1-\ell_2}\oplus Q_{n}$ by the similarity matrix $(\begin{smallmatrix}S&0\\ 0&T\end{smallmatrix})$ with $S=(\begin{smallmatrix}1&-x^{\ell_1}\\ 0&-1\end{smallmatrix})$ and $T=(\begin{smallmatrix}0&1\\ 1&x^{\ell_2}\end{smallmatrix})$, while in the case $\ell_1+\ell_2>n$ the cone of the element $(\begin{smallmatrix}x^{\ell_1+\ell_2-n}&0\\ 0&-1\end{smallmatrix})\in H^0(Q_{\ell_1},\bar{Q}_{\ell_2})$ can be transformed into $Q_{0}\oplus Q_{2n-\ell_1-\ell_2}$ by the similarity matrix $(\begin{smallmatrix}U&0\\ 0&V\end{smallmatrix})$ with $U=(\begin{smallmatrix}-1&-x^{n-\ell_2}\\ 0&1\end{smallmatrix})$ and $V=(\begin{smallmatrix}0&1\\ 1&x^{n-\ell_1}\end{smallmatrix})$. Thus the precise equivalent to the result of~\cite{h0401} is
\begin{align*}
&\text{C}\left[\begin{pmatrix}-1&0\\ 0&x^{n-\ell_1-\ell_2}\end{pmatrix}: Q_{\ell_1} \longrightarrow \bar{Q}_{\ell_2}\right] \cong \bar{Q}_{\ell_1+\ell_2} && \quad \text{for } \ell_1+\ell_2\leq n, \\
&\text{C}\left[\begin{pmatrix}x^{\ell_1+\ell_2-n}&0\\ 0&-1\end{pmatrix}: Q_{\ell_1} \longrightarrow \bar{Q}_{\ell_2}\right] \cong \bar{Q}_{\ell_1+\ell_2-n} && \quad \text{for } \ell_1+\ell_2 > n,
\end{align*}
where now the mapping cone has a direct interpretation as tachyon condensation. Furthermore, the fermionic state $(\begin{smallmatrix}0&x^{n-\ell_1-\ell_2}\\ -1&0\end{smallmatrix})\in H^1(Q_{\ell_1},Q_{\ell_2})$ corresponding to $(\begin{smallmatrix}-1&0\\ 0&x^{n-\ell_1-\ell_2}\end{smallmatrix})\in H^0(Q_{\ell_1},\bar{Q}_{\ell_2})$ in the case $\ell_1+\ell_2\leq n$ is directly related to the ``generator'' of Hori's flows, see~\cite[eq.\,(4.14)]{h0401}. Hence this fermionic state is really to be viewed as the tachyon that drives the condensation.

\vspace{0.2cm}

\noindent\textbf{Bulk RG flows. } One can also immediately obtain some of the details of bulk perturbations for minimal models in terms of matrix factorisations, i.\,e.~the effects of deformations of the Landau-Ginzburg potential $W_0=x^n$ of the form $W_\lambda=x^n+\sum_{i=m}^{n-1}\lambda_ix^i$ with $\lambda_m\neq 0$, say $\lambda_m=1$. Because of Orlov's equivalence~\cite{o0302,o0503} between the category of matrix factorisations $\text{MF}(W)$ and the category of singularities $\boldsymbol{D}_{\text{Sg}}(\C[X]/W)=\boldsymbol{D}^{\text{b}}(\text{mod--}\C[X]/W)/\text{Perf}(\C[X]/W)$, where the Verdier quotient is taken with respect to the category of perfect complexes whose objects are quasi-isomorphic to bounded complexes of projective modules, it follows that $\text{MF}(W_\lambda)\cong\text{MF}(x^m)$. The reason is that the perturbed potential can be factorised as $W_\lambda(x)=x^mf_\lambda(x)$ with $f_\lambda(x)=1+\sum_{i=m+1}^{n-1}\lambda_ix^{i-m}+x^{n-m}$ such that
$$
\C[X]/W_\lambda \cong \C[X]/(x^m)\oplus \C[X]/f_\lambda \, ,
$$
but $f_\lambda$ is not singular, and so after localisation the term $\C[X]/f_\lambda$ does not contribute to the category of singularities. 

This is consistent with the conjecture that matrix factorisations describe the open topological conformal field theory which is the infra-red fixed point of the RG flow of the Landau-Ginzburg model: In the IR limit, the lowest-order term $x^m$ in $W_\lambda$ is dominant, though the flow from the Landau-Ginzburg model to the conformal field theory may be complicated. But matrix factorisations only capture the situation at the fixed point, so as soon as perturbations with any $\lambda_i\neq 0$ are introduced, the formalism of matrix factorisations directly ``jumps'' to the new fixed point. Hence it is to be expected that $\text{MF}(W_\lambda)\cong\text{MF}(x^m)$ from the physical intuition as well. 

Although the equivalence of categories predicts that the topological theory at the IR fixed point is equivalently described by $\text{MF}(x^m)$, this fact alone does not provide the details of the RG flow. In particular, the theory with potential $x^n$ has $n-1$ fundamental D-branes, while the theory with the perturbed potential $x^n+\sum_{i=m}^{n-1}\lambda_ix^i$ has only $m-1$ fundamental D-branes. These $m-1$ branes can easily be identified together with those branes in $\text{MF}(x^n)$ which flow to them: There are $2m$ obvious matrix factorisations of $W_\lambda=x^n+\sum_{i=m}^{n-1}\lambda_ix^i$, namely
$$
Q'_i = \begin{pmatrix}0&W_\lambda/x^i\\x^i&0\end{pmatrix}
$$
together with their anti-branes for $i\in\{1,\ldots,m\}$. Now the claim is that only $m-1$ of these $2m$ branes are independent: Firstly, one can easily check that $H^0(Q'_m,Q'_m)=0$; in particular this means that there is no identity for the object $Q'_m$ and hence it must be isomorphic to the zero object.\footnote{Note that contrary to the impression sometimes given in the physics literature, this is \textit{not} the same as the existence of a unimodular polynomial matrix that similarity-transforms $Q'_m$ into the trivial matrix factorisation $(\begin{smallmatrix}0&W_\lambda\\1&0\end{smallmatrix})$ or $(\begin{smallmatrix}0&1\\ W_\lambda&0\end{smallmatrix})$; such a matrix simply does not exist in the present case. What is true is that $Q'_m$ is isomorphic to zero because its identity morphism in the associated DG category is BRST-exact and therefore zero in $\text{MF}(W_\lambda)$.} Secondly, for all $k\in\{1,\ldots,\lfloor\frac{m}{2}\rfloor\}$ one has $Q'_k\cong \bar{Q}'_{m-k}$. This can be proven by noting that for all such~$k$, $\phi_k=(\begin{smallmatrix}-1&0\\0&\sum_{i=0}^{n-m-1}\lambda_{m+i}x^i + x^{n-m}\end{smallmatrix})$ is an element of $H^0(Q'_k,\bar{Q}'_{m-k})$. But to say that $\phi_k$ is an isomorphism is the same as to say that the object $\text{C}(\phi_k)$ is isomorphic to zero. And indeed, using the Smith form once again one finds that $V\text{C}(\phi_k)V^{-1}$ is equal to
$$
\begin{pmatrix}0&0&x^m&0\\ 0&0&0&1\\ \sum_{i=0}^{n-m-1}\lambda_{m+i}x^i + x^{n-m}&0&0&0\\ 0&x^n+\sum_{i=m}^{n-1}\lambda_ix^i&0&0\end{pmatrix} \cong -\bar{Q}'_m\oplus 0\cong 0
$$
with
$$
V=\begin{pmatrix}1&-x^k&0&0\\ 0&-1&0&0\\ 0&0&0&1\\ 0&0&1&x^{m-k}\end{pmatrix} \, .
$$
A straightforward analysis of morphism spaces shows that none of the branes $Q'_1,\ldots,Q'_{\lfloor\frac{m}{2}\rfloor}$ and their anti-branes are pairwise isomorphic. Thus they form a complete set of fundamental branes in the perturbed theory $\text{MF}(W_\lambda)$. Furthermore, the branes $Q_i$ and $Q_{n-m+i}$ in $\text{MF}(x^n)$ both flow to the branes $Q'_i$ in $\text{MF}(W_\lambda)$ for $i\in\{1,\ldots,\lfloor\frac{m}{2}\rfloor\}$, and the respective anti-branes have the corresponding behaviour. 

On the other hand, the argument so far says nothing about the flow of the branes $Q_j$ in $\text{MF}(x^n)$ for $j\in\{m,\ldots,n-m\}$. It is expected that the bulk RG flow is described by a functor $\text{MF}(x^n)\rightarrow\text{MF}(W_\lambda)\cong\text{MF}(x^m)$, but its action on the remaining branes does not seem obvious from the matrix factorisations formalism alone. One possible way to solve this problem might be to try and solve the coupled bulk and boundary RG flow equations from perturbative conformal field theory (see e.\,g.~\cite{fgk0609034}) and then generalise the method of~\cite{rrs0003110} to bulk-induced boundary flows in supersymmetric theories. Much of the calculations of the involved operator product expansion coefficients may be simplified by the known correspondence between (topological) conformal field theory and matrix factorisations, however this direction will not be pursued further in the present paper.

\subsection{ADE-type D}\label{ADED}

In the study of tachyon condensation in Landau-Ginzburg models with potential $W_{\text{D}_\ell}=x^2y+y^{\ell-1}+z^2$ one cannot hope for the simplifications that arise in the single variable case as in the previous subsection. This offers an opportunity to apply the method of section~\ref{Preliminaries} to prove a similar generation result. 

It is known~\cite{schreyer1987,yoshino,kst0511} that the category $\text{MF}(W_{\text{D}_\ell})$ has a complete set of indecomposable objects $Q_i=(\begin{smallmatrix}0&g_i\\ f_i&0\end{smallmatrix})$ given by
\begin{align*}
& f_j = g_j = \begin{pmatrix}
-z & y^{\frac{j+1}{2}} & xy & 0 \\
y^{\ell-\frac{j+3}{2}} & z & 0 & -x \\
x & 0 & z & y^{\frac{j-1}{2}} \\
0 & -xy & y^{\ell-\frac{j+1}{2}} & -z
\end{pmatrix} && \text{for } j \text{ odd and } 1\leq j\leq\ell-2, \\
& f_j = g_j = \begin{pmatrix}
-z & 0 & xy & y^{\frac{j}{2}} \\
0 & -z & y^{\ell-1-\frac{j}{2}} & -x \\
x & y^{\frac{j}{2}} & z & 0 \\
y^{\ell-1-\frac{j}{2}} & -xy & 0 & z
\end{pmatrix} && \text{for } j \text{ even and } 2\leq j\leq\ell-2, \\
& f_{\ell-1} = g_{\ell-1} = \begin{pmatrix}
z & xy+\I y^{\frac{\ell}{2}} \\
x-\I y^{\frac{\ell-2}{2}} & -z
\end{pmatrix} && \text{for } \ell \text{ even,} \\
& f_{\ell} = g_{\ell} = \begin{pmatrix}
z & xy-\I y^{\frac{\ell}{2}} \\
x+\I y^{\frac{\ell-2}{2}} & -z
\end{pmatrix} && \text{for } \ell \text{ even,} \\
& f_{\ell-1} = g_{\ell} = \begin{pmatrix}
z+\I  y^{\frac{\ell-1}{2}} & xy \\
x & -z+\I y^{\frac{\ell-1}{2}}
\end{pmatrix} && \text{for } \ell \text{ odd,} \\
& f_{\ell} = g_{\ell-1} = \begin{pmatrix}
z-\I  y^{\frac{\ell-1}{2}} & xy \\
x & -z-\I y^{\frac{\ell-1}{2}}
\end{pmatrix} && \text{for } \ell \text{ odd.}
\end{align*}
Note that $Q_1$ has constant entries and can thus be reduced to a lower-rank matrix factorisation: it is isomorphic to $(\begin{smallmatrix}0&\phi\\ \phi&0\end{smallmatrix})$ with $\phi=(\begin{smallmatrix}z&x^2+y^{\ell-2}\\ y&-z\end{smallmatrix})$. 

The details of generating processes in $\text{MF}(W_{\text{D}_\ell})$ depend on whether $\ell$ is even or odd, and both cases will now be treated in turn.

\vspace{0.2cm}

\noindent\textbf{$\boldsymbol{\ell}$ even. } Let $\ell=2n$ with $n\geq 2$. Then the cones of the morphisms represented by
$$
\varphi_j = \text{diag}(y^j,y^j,y^j,y^j) \equiv \begin{pmatrix}y^j&0\\ 0&y^j\end{pmatrix} \oplus \begin{pmatrix}y^j&0\\ 0&y^j\end{pmatrix}
$$
in $H^0(Q_{2n},Q_{2n})$ are isomorphic to the branes $Q_{2j}$ for $j\in\{1,\ldots,n-1\}$. The corresponding isomorphisms in $H^0(Q_{2j},\text{C}(\varphi_j))$ are explicitly represented by
$$
\begin{pmatrix}
0&0&-y^{n-j}&\I\\
-y^{n-1-j}&-\I&0&0\\
-\I&0&0&0\\
0&0&\I&0
\end{pmatrix}
\oplus
\begin{pmatrix}
0&0&y^{n-j}&\I\\
-y^{n-1-j}&\I&0&0\\
\I&0&0&0\\
0&0&-\I&0
\end{pmatrix}
$$
as can be checked by verifying that the cones of the latter have empty endomorphism spaces. To generate also the fundamental branes labelled by odd indices, consider the morphisms represented by
$$
\psi_j = \begin{pmatrix}0&-y^j\\ y^{j-1}&0\end{pmatrix} \oplus \begin{pmatrix}0&y^j\\ -y^{j-1}&0\end{pmatrix}
$$
in $H^0(Q_{2n-1},Q_{2n})$ for $j\in\{1,\ldots,n-1\}$. Firstly, $Q_1$ is isomorphic to the cone of $\psi_1$ with the isomorphism represented by
$$
\begin{pmatrix}
0&y^{n-1}-\I x&0&0\\
\I&0&0&0\\
0&-\I&0&0\\
0&0&0&0
\end{pmatrix}
\oplus
\begin{pmatrix}
0&y^{n-1}-\I x&0&0\\
\I&0&0&0\\
0&\I&0&0\\
0&0&0&0
\end{pmatrix}
$$
in $H^0(Q_{1},\text{C}(\psi_1))$, where the peculiar appearance of this isomorphism is due to the fact mentioned earlier that $Q_1$ can be reduced to a rank~2 matrix factorisation. Secondly, the matrices
$$
\begin{pmatrix}
-y^{n-j}&0&0&-\I\\
0&-\I&y^{n-j}&0\\
\I&0&0&0\\
0&0&-\I&0
\end{pmatrix}
\oplus
\begin{pmatrix}
-y^{n-j}&0&0&-\I\\
0&-\I&y^{n-j}&0\\
-\I&0&0&0\\
0&0&\I&0
\end{pmatrix}
$$
represent isomorphisms in $H^0(Q_{2j-1},\text{C}(\psi_j))$ for all $j\in\{2,\ldots,n-1\}$.

The above calculations show that with two types of fundamental branes, $Q_{2n}$ and $Q_{2n-1}$, every D-brane system can be generated in Landau-Ginzburg models of type D$_{2n}$. One may wonder whether it is possible to find further tachyon condensations such that actually only one fundamental brane is needed, as for example in models of type~A. But according to table~\ref{ADEGrothendieck} the associated Grothendieck group is $\Z_2\oplus\Z_2$ in the present case. Thus one brane alone cannot account for all possible charges, which shows that the two branes identified above indeed represent a minimal configuration: $\text{MF}(W_{\text{D}_{2n}})=\text{tria}(Q_{2n-1},Q_{2n})$.

On the other hand, $Q_{2n}$ and $Q_{2n-1}$ are not the only pair of branes with this generation property. Computations completely analogous to the above show that $\text{MF}(W_{\text{D}_{2n}})$ is also generated by $Q_{2n}$ and $Q_{2j-1}$, or by $Q_{2n-1}$ and $Q_{2j-1}$ for all $j\in\{2,\ldots,n-1\}$. What is not possible is to generate the D-brane category only from oddly-labelled branes $Q_{2j-1}$, or only from an evenly-labelled brane $Q_{2j}$ and one additional arbitrary brane. The reason for this is that the Grothendieck group $K_0(\text{MF}(W_{\text{D}_{2n}}))$ is a non-trivial direct sum, but for all oddly-labelled branes one has $Q_{2j-1}\cong\text{C}(\psi_j)$ with $\psi_j\in H^0(Q_{2n-1},Q_{2n})$, and hence distinguished triangles
$$
Q_{2n-1}\longrightarrow Q_{2n}\longrightarrow Q_{2j-1}\longrightarrow \bar Q_{2n-1} \, .
$$
Consequently all the branes $Q_{2j-1}$ with $j\in\{2,\ldots,n-1\}$ by definition represent the same element in the Grothendieck group, and thus cannot generate the whole category. Similarly, the evenly-labelled branes $Q_{2j}\cong\text{C}(\varphi_j)$ represent the zero element as they are isomorphic to cones of endomorphisms $\varphi_j\in H^0(Q_{2n},Q_{2n})$.

The preceeding arguments exemplify the advantages of using the structure of matrix factorisations over analyses in conformal field theory in certain situations: while the computation of Grothendieck groups is rather straightforward for Landau-Ginzburg models of type ADE, there is currently no general method available to compute torsion charges~\cite{bd0102018,bdm0106262} in the associated $N=2$ minimal superconformal field theories. In particular, it is not known how to extract torsion charges by computing overlaps with the D-model boundary states of~\cite{bg0506} corresponding to the branes analysed here.

\vspace{0.2cm}

\noindent\textbf{$\boldsymbol{\ell}$ odd. } Let $\ell=2n+1$ with $n\geq 2$. Then the cones of the morphisms represented by
$$
\varphi_j = \begin{pmatrix}y^j&0\\ 0&y^j\end{pmatrix} \oplus \begin{pmatrix}y^j&0\\ 0&y^j\end{pmatrix}
$$
in $H^0(Q_{2n+1},Q_{2n+1})$ are isomorphic to the branes $Q_{2j}$ for $j\in\{1,\ldots,n-1\}$. The corresponding isomorphisms in $H^0(Q_{2j},\text{C}(\varphi_j))$ are represented by
$$
\begin{pmatrix}
-y^{n-j}&0&0&-\I\\
0&\I&y^{n-j}&0&\\
\I&0&0&0\\
0&0&-\I&0
\end{pmatrix}
\oplus
\begin{pmatrix}
-y^{n-j}&0&0&\I\\
0&-\I&y^{n-j}&0&\\
-\I&0&0&0\\
0&0&\I&0
\end{pmatrix} \, .
$$
As in the previous case of even $\ell$, the cones of morphisms between $Q_{\ell-1}$ and $Q_\ell$ can be shown to be isomorphic to the fundamental branes labelled by odd indices. But in contrast to the previous case, $Q_{\ell-1}$ can now be understood as a non-trivial composite object: it is isomorphic to the cone of the morphism represented by
$$
\xi_n=
\begin{pmatrix}
-\I & 0 &   0 & 0 \\
0 &    -y^n & 0 & 0 \\ 
0 &    \I & 0 & 0 \\
y^n &   0 &   0 & 0
\end{pmatrix}
\oplus
\begin{pmatrix}
\I & 0 &    0 & 0 \\
0 &   -y^n &  0 & 0 \\
0 &   -\I & 0 & 0 \\
y^n &  0 &    0 & 0
\end{pmatrix}
$$
in $H^0(Q_{2n+1},Q_2)$, where the isomorphism is represented by
$$
\begin{pmatrix}
-y &  0 &    0 & 0 & 0 & 0 & 0 & 0 \\
0 &   -y &   0 & 0 & 0 & 0 & 0 & 0 \\
0 &   0 &    0 & 0 & 0 & 0 & 0 & 0 \\
0 &   0 &    0 & 0 & 0 & 0 & 0 & 0 \\
0 &   0 &    0 & 0 & 0 & 0 & 0 & 0 \\
0 &   -\I & 0 & 0 & 0 & 0 & 0 & 0 \\
0 &   0 &    0 & 0 & 0 & 0 & 0 & 0\\ 
\I & 0 &    0 & 0 & 0 & 0 & 0 & 0
\end{pmatrix}
\oplus
\begin{pmatrix}
y &   0 &    0 & 0 & 0 & 0 & 0 & 0 \\
0 &   y &    0 & 0 & 0 & 0 & 0 & 0 \\
0 &   0 &    0 & 0 & 0 & 0 & 0 & 0 \\
0 &   0 &    0 & 0 & 0 & 0 & 0 & 0 \\
0 &   0 &    0 & 0 & 0 & 0 & 0 & 0 \\
0 &   -\I & 0 & 0 & 0 & 0 & 0 & 0 \\
0 &   0 &    0 & 0 & 0 & 0 & 0 & 0 \\
\I & 0 &    0 & 0 & 0 & 0 & 0 & 0
\end{pmatrix}
$$
in $H^0(Q_{2n},\text{C}(\xi_n))$. Now consider the morphisms represented by
$$
\psi_j = \begin{pmatrix}0&-y^j\\ y^{j-1}&0\end{pmatrix} \oplus \begin{pmatrix}0&y^j\\ -y^{j-1}&0\end{pmatrix}
$$
in $H^0(Q_{2n},Q_{2n+1})$ for $j\in\{1,\ldots,n\}$. Firstly, $Q_1$ is isomorphic to the cone of $\psi_1$ with the isomorphism represented by
$$
\begin{pmatrix}
0&x&0&0\\
-1&-\I y^2&0&0\\
0&1&0&0\\
0&0&0&0
\end{pmatrix}
\oplus
\begin{pmatrix}
0&x&0&0\\
-1&\I y^2&0&0\\
0&-1&0&0\\
0&0&0&0
\end{pmatrix}
$$
in $H^0(Q_{1},\text{C}(\psi_1))$. Secondly, the matrices
$$
\begin{pmatrix}
0&0&-y^{n+1-j}&-\I\\
-y^{n-j}&-\I&0&0\\
\I&0&0&0\\
0&0&-\I&0
\end{pmatrix}
\oplus
\begin{pmatrix}
0&0&y^{n+1-j}&-\I\\
y^{n-j}&-\I&0&0\\
-\I&0&0&0\\
0&0&\I&0
\end{pmatrix}
$$
represent isomorphisms in $H^0(Q_{2j-1},\text{C}(\psi_j))$ for all $j\in\{2,\ldots,n\}$.

In summary, every D-brane in $\text{MF}(W_{\text{D}_{2n+1}})$ can be viewed as a (repeated) tachyon condensation of the single brane $Q_{2n+1}$: $\text{MF}(W_{\text{D}_{2n+1}})=\text{tria}(Q_{2n+1})$. But the brane $Q_{2n+1}$ is not the only one with this property: by computations very similar to the above one can show that $Q_{2n}$ may also serve as a single generator of $\text{MF}(W_{\text{D}_{2n+1}})$. On the other hand, all remaining fundamental branes $Q_i$ fail to generate the whole D-brane category. For the evenly-labelled $Q_{2j}$, this is again due to the fact that they are isomorphic to cones of endomorphisms and thus represent the zero element in the Grothendieck group $K_0(\text{MF}(W_{\text{D}_{2n+1}}))=\Z_4$. It follows from the above results that $Q_{2n+1}$ and $Q_{2n}$ have ``opposite'' charges in this group, and because of their generation property they must therefore represent the elements 1 and 3 in $\Z_4$. Now since the oddly-labelled branes $Q_{2j-1}$ are isomorphic to cones of morphisms between $Q_{2n}$ and $Q_{2n+1}$, it is clear that $[Q_{2j-1}]=2\in\Z_4$. This proves that only $Q_{2n+1}$ and $Q_{2n}$ can triangle-generate $\text{MF}(W_{\text{D}_{2n+1}})$ by themselves. 

\subsection{ADE-type E}\label{ADEE}

The same strategy of analysing double-cones will now be applied to the three individual models of ADE-types E$_6$, E$_7$ and E$_8$. The result is that the corresponding categories are each triangle-generated by a single indecomposable object, meaning that all D-branes in the Landau-Ginzburg model are tachyon condensates of the associated fundamental brane. 

This generation property can be shown by explicitly describing how all $n$ indecomposable objects of $\text{MF}(W_{\text{E}_n})$ for $n\in\{6,7,8\}$ can be viewed as (repeated) cones pertaining to $Q_n$. The details of these computations are presented in the appendix, and the results can be summarised by
\begin{align*}
\text{E}_6: & \quad Q_6,Q_6\rightsquigarrow Q_1 \, , \quad Q_1,Q_1\rightsquigarrow Q_2 \, , \quad Q_1,\bar Q_6\rightsquigarrow Q_4 \, , \quad Q_1,\bar Q_6\rightsquigarrow Q_5 \, , \\
& \quad Q_1,Q_6\rightsquigarrow Q_3 \, , \\
\text{E}_7: & \quad Q_7,Q_7\rightsquigarrow Q_1 \, , \quad Q_7,Q_7\rightsquigarrow Q_6 \, , \quad Q_1,Q_1\rightsquigarrow Q_2 \, , \quad Q_6,Q_6\rightsquigarrow Q_3 \, , \\
& \quad Q_1,Q_7\rightsquigarrow Q_4 \, , \quad Q_1,Q_4\rightsquigarrow Q_5 \, , \\
\text{E}_8: & \quad Q_8,Q_8\rightsquigarrow Q_1 \, , \quad Q_8,Q_8\rightsquigarrow Q_3 \, , \quad Q_8,Q_8\rightsquigarrow Q_6 \, , \quad Q_8,Q_8\rightsquigarrow Q_7 \, , \\
& \quad Q_1,Q_1\rightsquigarrow Q_2 \, , \quad Q_1,Q_3\rightsquigarrow Q_4 \, , \quad Q_6,Q_6\rightsquigarrow Q_5 \, ,
\end{align*}
where $Q_i, Q_j\rightsquigarrow Q_k$ should be read as ``the indecomposable object $Q_k$ is isomorphic to the cone of a morphism in $H^0(Q_i,Q_j)$''. By choosing basis elements for the $n^2$ cohomologies between $Q_i$ and $Q_j$ for all $i,j\in\{1,\ldots,n\}$, one can of course also compute all of their respective cones to get a more complete picture of tachyon condensation in the E$_n$ models. A sketch of the results for E$_6$ is displayed in table~\ref{E6cones}. 

\begin{table}[t]
\begin{center}
\begin{tabular}{@{}*{1}{l}*{6}{r}@{}}
\toprule[2pt]
& $j=1$ & $j=2$ & $j=3$ & $j=4$ & $j=5$ & $j=6$
\tabularnewline\midrule[1.5pt]
$i=1$ & $Q_2$ & & $Q_6$ & $Q_5$ & $Q_4$ & $Q_3$ \tabularnewline
\addlinespace
$i=2$ &  & $Q_1$ & $Q_6$ & $Q_5$ & $Q_4$ & $Q_3$ \tabularnewline
\addlinespace
$i=3$ & $Q_5$ & $Q_4, Q_5$ & & $Q_6$ & $Q_6$ & $Q_1, Q_2$ \tabularnewline
\addlinespace
$i=4$ & $Q_6$ & $Q_3, Q_6$ & $Q_5$ &  & $Q_1, Q_2$ & $Q_5$ \tabularnewline
\addlinespace
$i=5$ & $Q_3, Q_6$ & $Q_3$ & $Q_4$ & $Q_1, Q_2$ & $Q_1$ & $Q_4$ \tabularnewline
\addlinespace
$i=6$ & $Q_4, Q_5$ & $Q_4$ & $Q_1, Q_2$ & $Q_3$ & $Q_3$ & $Q_1$ \tabularnewline
\bottomrule[2pt]
\end{tabular}%
\caption{Cones of $Q_i\rightarrow Q_j$ in $\text{MF}(W_{\text{E}_6})$.\label{E6cones}}%
\end{center}%
\end{table}%

\section*{}

\subsection*{Conclusions}

The main idea of this note is that one can systematically and algorithmically analyse tachyon condensation in triangulated D-brane categories if one has a good handle on the explicit computation of morphism spaces. For the category of matrix factorisations a rather simple and general algorithm for these computations was described and tested for all Landau-Ginzburg models of type ADE. 

An immediate next step would of course be to apply this method to more complicated models and study the relations between their various types of D-branes. For example, one may look more closely at the linear matrix factorisations of~\cite{err0508}. All known boundary states of the conformal field theories which correspond to the Landau-Ginzburg models where linear matrix factorisations arise can be mapped to such factorisations. But there are also linear matrix factorisations that have no interpretation in terms of conformal field theory or geometry so far. It would be interesting to understand which of these ``new'' topological branes can be viewed as condensates of ``old'' ones. 

Furthermore, the work of~\cite{keller1994,l0310337,l0610120} suggests that knowledge of triangle-generation properties of $\text{MF}(W)$ can also be helpful to understand matrix factorisations to have the full structure of open topological string theories, i.\,e.~of cyclic, unital and minimal $\Ainf$-categories. This approach should be studied further.

\vspace{0.2cm}

\noindent\textbf{Acknowledgements. } I thank P.~S.~Aspinwall, M.~Baumgartl, M.~R.~Gaberdiel, H.~Kajiura, C.~I.~Lazaroiu, D.~Orlov, A.~Quintero V\'{e}lez and especially A.~Recknagel for discussions and correspondence. For their support I am grateful to R.~Streater, A.~Wi\ss kirchen, Deutscher Akademischer Auslandsdienst, School of Physical Sciences and Engineering of King's College London, and Studienstiftung des deutschen Volkes.

\appendix

\section{Explicit results for models of type E}

\subsection*{ADE-type $\boldsymbol{\text{E}_6}$}

It is known~\cite{schreyer1987,yoshino,kst0511} that the category $\text{MF}(W_{\text{E}_6})$ with $W_{\text{E}_6}=x^3+y^4+z^2$ has a complete set of indecomposable objects $Q_i=(\begin{smallmatrix}0&g_i\\ f_i&0\end{smallmatrix})$ given by
\begin{align*}
f_1 = & \begin{pmatrix}
-z & 0 & x^{2} & y^{3} \\
0 & -z & y & -x \\
x & y^{3} & z & 0 \\
y & -x^{2} & 0 & z
\end{pmatrix} \, , \quad
g_1 = \begin{pmatrix}
-z & 0 & x^{2} & y^{3} \\
0 & -z & y & -x \\
x & y^{3} & z & 0 \\
y & -x^{2} & 0 & z
\end{pmatrix} \, , \\
f_2 = & \begin{pmatrix}
-\I z & -y^{2} & xy & 0 & x^{2} & 0 \\
-y^{2} & -\I z & 0 & 0 & 0 & x \\
0 & 0 & -\I z & -x & 0 & y \\
0 & xy & -x^{2} & -\I z & y^{3} & 0 \\
x & 0 & 0 & y & -\I z & 0 \\
0 & x^{2} & y^{3} & 0 & xy^{2} & -\I z
\end{pmatrix} \, , \\
g_2 = & \begin{pmatrix}
\I z & -y^{2} & xy & 0 & x^{2} & 0 \\
-y^{2} & \I z & 0 & 0 & 0 & x \\
0 & 0 & \I z & -x & 0 & y \\
0 & xy & -x^{2} & \I z & y^{3} & 0 \\
x & 0 & 0 & y & \I z & 0 \\
0 & x^{2} & y^{3} & 0 & xy^{2} & \I z
\end{pmatrix} \, , \\
f_3 = & \begin{pmatrix}
-y^{2}+\I z & 0 & xy & x \\
-xy & y^{2}+\I z & x^{2} & 0 \\
0 & x & \I z & y \\
x^{2} & -xy & y^{3} & \I z
\end{pmatrix} \, , \quad
g_3 = \begin{pmatrix}
-y^{2}-\I z & 0 & xy & x \\
-xy & y^{2}-\I z & x^{2} & 0 \\
0 & x & -\I z & y \\
x^{2} & -xy & y^{3} & -\I z
\end{pmatrix} \, , \\
f_4 =& \begin{pmatrix}
-y^{2}-\I z & 0 & xy & x \\
-xy & y^{2}-\I z & x^{2} & 0 \\
0 & x & -\I z & y \\
x^{2} & -xy & y^{3} & -\I z
\end{pmatrix} \, , \quad 
g_4 = \begin{pmatrix}
-y^{2}+\I z & 0 & xy & x \\
-xy & y^{2}+\I z & x^{2} & 0 \\
0 & x & \I z & y \\
x^{2} & -xy & y^{3} & \I z
\end{pmatrix} \, , \\
f_5 = & \begin{pmatrix}
-y^{2}+\I z & x \\
x^{2} & y^{2}+\I z
\end{pmatrix} \, , \quad
g_5 = \begin{pmatrix}
-y^{2}-\I z & x \\
x^{2} & y^{2}-\I z
\end{pmatrix} \, , \\
f_6 = & \begin{pmatrix}
-y^{2}-\I z & x \\
x^{2} & y^{2}-\I z
\end{pmatrix} \, , \quad
g_6 = \begin{pmatrix}
-y^{2}+\I z & x \\
x^{2} & y^{2}+\I z
\end{pmatrix} \, .
\end{align*}
$\text{MF}(W_{\text{E}_6})$ is generated by $Q_6$. To see this, first note that $\varphi_1=\text{diag}(y,y,y,y)=(\begin{smallmatrix}y&0\\ 0&y\end{smallmatrix})\oplus(\begin{smallmatrix}y&0\\ 0&y\end{smallmatrix})$ represents an element in the 2-dimensional space $H^0(Q_6,Q_6)$, and the element represented by
$$
\begin{pmatrix}
0 &   -\I y & 1 & 0 \\
\I  & 0 &      0 & y \\
0 &   -\I  &   0 & 0 \\
0 &   0 &      0 & -1
\end{pmatrix}
\oplus
\begin{pmatrix}
0 &  y & -\I  & 0 \\
-1 & 0 & 0 &    -\I y \\
0 &  1 & 0 &    0 \\
0 &  0 & 0 &    \I 
\end{pmatrix}
$$
in the 4-dimensional space $H^0(Q_1,\text{C}(\varphi_1))$ has a zero-isomorphic cone. Hence $Q_1$ is generated by $Q_6$. Next, 
$$
\varphi_2=
\begin{pmatrix}
0 &  xy^2 & 0 & 0 \\
-1 & 0 &   0 & 0 \\
0 &  0 &   0 & -y^2 \\
0 &  0 &   x & 0
\end{pmatrix}
\oplus
\begin{pmatrix}
0 &  xy^2 & 0 & 0 \\
-1 & 0 &   0 & 0 \\
0 &  0 &   0 & -y^2 \\
0 &  0 &   x & 0
\end{pmatrix}
$$
represents an element in the 4-dimensional space $H^0(Q_1,Q_1)$, and the element represented by
$$
\begin{pmatrix}
0 &  \I y & -\I x & 0 & 0 &   0 &  0 & 0 \\
0 &  0 &     0 &      0 & \I  & 0 &  0 & 0 \\
0 &  0 &     0 &      1 & 0 &   0 &  0 & 0 \\
-1 & 0 &     0 &      0 & 0 &   0 &  0 & 0 \\
0 &  0 &     0 &      0 & 0 &   -1 & 0 & 0 \\
0 &  0 &     0 &      0 & 0 &   0 &  0 & 0 \\
0 &  \I  &   0 &      0 & 0 &   0 &  0 & 0 \\
0 &  0 &     \I  &    0 & 0 &   0 &  0 & 0
\end{pmatrix}
\oplus
\begin{pmatrix}
0 &   -y & x &  0 &    0 &  0 &   0 & 0 \\
0 &   0 &  0 &  0 &    -1 & 0 &   0 & 0 \\
0 &   0 &  0 &  -\I  & 0 &  0 &   0 & 0 \\
\I  & 0 &  0 &  0 &    0 &  0 &   0 & 0 \\
0 &   0 &  0 &  0 &    0 &  \I  & 0 & 0 \\
0 &   0 &  0 &  0 &    0 &  0 &   0 & 0 \\
0 &   -1 & 0 &  0 &    0 &  0 &   0 & 0 \\
0 &   0 &  -1 & 0 &    0 &  0 &   0 & 0
\end{pmatrix}
$$
in the 12-dimensional space $H^0(Q_2,\text{C}(\varphi_2))$ has a zero-isomorphic cone. Hence $Q_2$ is generated by $Q_1$. Next, 
$$
\varphi_3=
\begin{pmatrix}
-1 & 0 &   0 &      \I y \\
0 &  -xy & -\I x & 0 \\
0 &  0 &   0 &      0 \\
0 &  0 &   0 &      0
\end{pmatrix}
\oplus
\begin{pmatrix}
-\I  & 0 &       0 &  y \\
0 &    -\I xy & -x & 0 \\
0 &    0 &       0 &  0 \\
0 &    0 &       0 &  0
\end{pmatrix}
$$
represents an element in the 2-dimensional space $H^0(Q_1,Q_6)$, and the element represented by
$$
\begin{pmatrix}
x &   -y &   0 & 0 &   0 & 0 & 0 & 0 \\
0 &   0 &    1 & 0 &   0 & 0 & 0 & 0 \\
0 &   0 &    0 & \I  & 0 & 0 & 0 & 0 \\
0 &   -\I  & 0 & 0 &   0 & 0 & 0 & 0 \\
0 &   0 &    0 & 0 &   0 & 0 & 0 & 0 \\
\I  & 0 &    0 & 0 &   0 & 0 & 0 & 0 \\
0 &   0 &    0 & 0 &   0 & 0 & 0 & 0 \\
0 &   0 &    0 & 0 &   0 & 0 & 0 & 0
\end{pmatrix}
\oplus
\begin{pmatrix}
-\I x & \I y & 0 &    0 &  0 & 0 & 0 & 0 \\
0 &      0 &     -\I  & 0 &  0 & 0 & 0 & 0 \\
0 &      0 &     0 &    -1 & 0 & 0 & 0 & 0 \\
0 &      1 &     0 &    0 &  0 & 0 & 0 & 0 \\
0 &      0 &     0 &    0 &  0 & 0 & 0 & 0 \\
-\I  &   0 &     0 &    0 &  0 & 0 & 0 & 0 \\
0 &      0 &     0 &    0 &  0 & 0 & 0 & 0 \\
0 &      0 &     0 &    0 &  0 & 0 & 0 & 0
\end{pmatrix}
$$
in the 6-dimensional space $H^0(Q_3,\text{C}(\varphi_3))$ has a zero-isomorphic cone. Hence $Q_3$ is generated by $Q_1$ and $Q_6$. Next, 
$$
\varphi_4=
\begin{pmatrix}
1 & 0 &  0 &      \I y \\
0 & xy & -\I x & 0 \\
0 & 0 &  0 &      0 \\
0 & 0 &  0 &      0
\end{pmatrix}
\oplus
\begin{pmatrix}
\I  & 0 &      0 &  y \\
0 &   \I xy & -x & 0 \\
0 &   0 &      0 &  0 \\
0 &   0 &      0 &  0
\end{pmatrix}
$$
represents an element in the 2-dimensional space $H^0(Q_1,\bar Q_6)$, and the element represented by
$$
\begin{pmatrix}
x &   -y &  0 & 0 &    0 & 0 & 0 & 0 \\
0 &   0 &   1 & 0 &    0 & 0 & 0 & 0 \\
0 &   0 &   0 & -\I  & 0 & 0 & 0 & 0 \\
0 &   \I  & 0 & 0 &    0 & 0 & 0 & 0 \\
0 &   0 &   0 & 0 &    0 & 0 & 0 & 0 \\
\I  & 0 &   0 & 0 &    0 & 0 & 0 & 0 \\
0 &   0 &   0 & 0 &    0 & 0 & 0 & 0 \\
0 &   0 &   0 & 0 &    0 & 0 & 0 & 0
\end{pmatrix}
\oplus
\begin{pmatrix}
\I x & -\I y & 0 &   0 &  0 & 0 & 0 & 0 \\
0 &     0 &      \I  & 0 &  0 & 0 & 0 & 0 \\
0 &     0 &      0 &   -1 & 0 & 0 & 0 & 0 \\
0 &     1 &      0 &   0 &  0 & 0 & 0 & 0 \\
0 &     0 &      0 &   0 &  0 & 0 & 0 & 0 \\
\I  &   0 &      0 &   0 &  0 & 0 & 0 & 0 \\
0 &     0 &      0 &   0 &  0 & 0 & 0 & 0 \\
0 &     0 &      0 &   0 &  0 & 0 & 0 & 0
\end{pmatrix}
$$
in the 6-dimensional space $H^0(Q_4,\text{C}(\varphi_4))$ has a zero-isomorphic cone. Hence $Q_4$ is generated by $Q_1$ and $Q_6$. Finally, 
$$
\varphi_5=
\begin{pmatrix}
0 &   -\I y & 1 & 0 \\
\I  & 0 &      0 & y \\
0 &   0 &      0 & 0 \\
0 &   0 &      0 & 0
\end{pmatrix}
\oplus
\begin{pmatrix}
0 &  y & -\I  & 0 \\
-1 & 0 & 0 &    -\I y \\
0 &  0 & 0 &    0 \\
0 &  0 & 0 &    0
\end{pmatrix}
$$
represents another element in the 2-dimensional space $H^0(Q_1,\bar{Q}_6)$, and the element represented by
$$
\begin{pmatrix}
0 &      -y &   0 & 0 & 0 & 0 & 0 & 0 \\
1 &      0 &    0 & 0 & 0 & 0 & 0 & 0 \\
-\I y & 0 &    0 & 0 & 0 & 0 & 0 & 0 \\
0 &      -\I  & 0 & 0 & 0 & 0 & 0 & 0 \\
0 &      0 &    0 & 0 & 0 & 0 & 0 & 0 \\
0 &      0 &    0 & 0 & 0 & 0 & 0 & 0 \\
0 &      0 &    0 & 0 & 0 & 0 & 0 & 0 \\
0 &      0 &    0 & 0 & 0 & 0 & 0 & 0
\end{pmatrix}
\oplus
\begin{pmatrix}
0 &    \I y & 0 & 0 & 0 & 0 & 0 & 0 \\
-\I  & 0 &     0 & 0 & 0 & 0 & 0 & 0 \\
y &    0 &     0 & 0 & 0 & 0 & 0 & 0 \\
0 &    1 &     0 & 0 & 0 & 0 & 0 & 0 \\
0 &    0 &     0 & 0 & 0 & 0 & 0 & 0 \\
0 &    0 &     0 & 0 & 0 & 0 & 0 & 0 \\
0 &    0 &     0 & 0 & 0 & 0 & 0 & 0 \\
0 &    0 &     0 & 0 & 0 & 0 & 0 & 0
\end{pmatrix}
$$
in the 2-dimensional space $H^0(Q_5,\text{C}(\varphi_5))$ has a zero-isomorphic cone. Hence $Q_5$ is generated by $Q_1$ and $Q_6$. 

In summary, every D-brane in $\text{MF}(W_{\text{E}_6})$ can be viewed as a (repeated) tachyon condensation of the single brane $Q_6$: $\text{MF}(W_{\text{E}_6})=\text{tria}(Q_6)$.

\subsection*{ADE-type $\boldsymbol{\text{E}_7}$}

The category $\text{MF}(W_{\text{E}_7})$ with $W_{\text{E}_7}=x^3+xy^3+z^2$ has a complete set of indecomposable objects $Q_i=(\begin{smallmatrix}0&f_i\\ f_i&0\end{smallmatrix})$ given by
\begin{align*}
f_1
= &
\begin{pmatrix}
z & 0 & -x^{2} & y \\
0 & z & xy^{2} & x \\
-x & y & -z & 0 \\
xy^{2} & x^{2} & 0 & -z
\end{pmatrix} \, , \\
f_2
= &
\begin{pmatrix}
-z & y^{2} & xy & 0 & x^{2} & 0 \\
xy & z & 0 & 0 & 0 & -x \\
0 & 0 & z & -x & 0 & y \\
0 & -xy & -x^{2} & -z & xy^{2} & 0 \\
x & 0 & 0 & y & z & 0 \\
0 & -x^{2} & xy^{2} & 0 & x^{2}y & -z
\end{pmatrix} \, , \\
f_3
= &
\begin{pmatrix}
-z & 0 & xy & -y^{2} & 0 & 0 & x^{2} & 0 \\
0 & -z & 0 & y^{2} & 0 & 0 & 0 & x \\
y^{2} & y^{2} & z & 0 & 0 & -x & 0 & 0 \\
0 & xy & 0 & z & -x^{2} & 0 & 0 & 0 \\
0 & 0 & 0 & -x & -z & 0 & 0 & y \\
0 & 0 & -x^{2} & 0 & 0 & -z & xy^{2} & y^{2} \\
x & 0 & 0 & 0 & -y^{2} & y & z & 0 \\
0 & x^{2} & 0 & 0 & xy^{2} & 0 & 0 & z
\end{pmatrix} \, , \\
f_4
= &
\begin{pmatrix}
-z & y^{2} & 0 & x \\
xy & z & -x^{2} & 0 \\
0 & -x & -z & y \\
x^{2} & 0 & xy^{2} & z
\end{pmatrix} \, , \\
f_5
= &
\begin{pmatrix}
-z & 0 & xy & 0 & 0 & x \\
-xy & z & 0 & -y^{2} & -x^{2} & 0 \\
y^{2} & 0 & z & -x & xy & 0 \\
0 & -xy & -x^{2} & -z & 0 & 0 \\
0 & -x & 0 & 0 & -z & -y \\
x^{2} & 0 & 0 & xy & -xy^{2} & z
\end{pmatrix} \, , \\
f_6
= &
\begin{pmatrix}
z & 0 & -xy & x \\
0 & z & x^{2} & y^{2} \\
-y^{2} & x & -z & 0 \\
x^{2} & xy & 0 & -z
\end{pmatrix} \, , \\
f_7
= &
\begin{pmatrix}
z & x \\
y^{3}+x^{2} & -z
\end{pmatrix} \, .
\end{align*}
$\text{MF}(W_{\text{E}_7})$ is generated by $Q_7$. To see this, first note that $\varphi_1=(\begin{smallmatrix}y&0\\ 0&y\end{smallmatrix})\oplus(\begin{smallmatrix}y&0\\ 0&y\end{smallmatrix})$ represents an element in the 3-dimensional space $H^0(Q_7,Q_7)$, and the element represented by
$$
\begin{pmatrix}0&-1&0&0\\ 0&0&-y^2&-1\\ 0&0&1&0\\ 1&0&0&0\end{pmatrix}
\oplus
\begin{pmatrix}0&1&0&0\\ 0&0&y^2&1\\ 0&0&-1&0\\ -1&0&0&0\end{pmatrix}
$$
in the 4-dimensional space $H^0(Q_1,\text{C}(\varphi_1))$ has a zero-isomorphic cone. Hence $Q_1$ is generated by $Q_7$. Next, $\varphi_2=(\begin{smallmatrix}y^2&0\\ 0&y^2\end{smallmatrix})\oplus(\begin{smallmatrix}y^2&0\\ 0&y^2\end{smallmatrix})$ represents another element in the 3-dimensional space $H^0(Q_7,Q_7)$, and the element represented by
$$
\begin{pmatrix}1&0&0&0\\ 0&0&-y&1\\ 0&0&1&0\\ 0&-1&0&0\end{pmatrix}
\oplus
\begin{pmatrix}-1&0&0&0\\ 0&0&y&-1\\ 0&0&-1&0\\ 0&1&0&0\end{pmatrix}
$$
in the 8-dimensional space $H^0(Q_6,\text{C}(\varphi_2))$ has a zero-isomorphic cone. Hence $Q_6$ is generated by $Q_7$. Next, 
$$
\varphi_3=
\begin{pmatrix}0&-x&0&0\\ 0&y^2&-yz&0\\ 0&0&0&1\\ -yz&0&0&-y^2\end{pmatrix}
\oplus
\begin{pmatrix}-y^2&-x&0&0\\ 0&0&yz&0\\ 0&0&y^2&1\\ yz&0&0&0\end{pmatrix}
$$
represents an element in the 4-dimensional space $H^0(Q_1,Q_1)$, and the element represented by
$$
\begin{pmatrix}-1&0&0&0&0&0&0&0\\ 0&0&0&1&0&0&0&0\\ 0&0&0&0&-1&0&0&0\\ 0&y&x&0&0&0&0&0\\ 0&0&1&0&0&0&0&0\\ 0&-1&0&0&y&0&0&0\\ 0&0&0&0&0&0&0&0\\ -y&0&0&0&0&1&0&0\end{pmatrix}
\oplus
\begin{pmatrix}-1&0&0&0&0&0&0&0\\ 0&0&0&1&0&0&0&0\\ 0&0&0&0&-1&0&0&0\\ 0&y&x&0&0&0&0&0\\ 0&0&1&0&0&0&0&0\\ 0&-1&0&0&0&0&0&0\\ 0&0&0&0&0&0&0&0\\ 0&0&0&0&0&1&0&0\end{pmatrix}
$$
in the 12-dimensional space $H^0(Q_2,\text{C}(\varphi_3))$ has a zero-isomorphic cone. Hence $Q_2$ is generated by $Q_1$. Next, 
$$
\varphi_4=
\begin{pmatrix}0&0&0&y\\ 0&0&-xy&0\\ 0&y&0&0\\ -xy&0&0&0\end{pmatrix}
\oplus
\begin{pmatrix}0&0&0&-y\\ 0&0&xy&0\\ 0&-y&0&0\\ xy&0&0&0\end{pmatrix}
$$
represents an element in the 8-dimensional space $H^0(Q_6,Q_6)$, and the element represented by
$$
\begin{pmatrix}-1&-1&0&0&0&0&0&0\\ 0&0&0&0&0&-1&0&0\\ 0&0&-1&0&0&0&0&0\\ 0&0&0&0&0&0&x&1\\ 0&0&0&0&1&0&0&0\\ 1&0&0&0&0&0&0&0\\ 0&0&0&0&0&0&-1&0\\ 0&0&0&1&0&0&0&0\end{pmatrix}
\oplus
\begin{pmatrix}-1&-1&0&0&0&0&0&0\\ 0&0&0&0&0&-1&0&0\\ 0&0&-1&0&0&0&0&0\\ 0&0&0&0&0&0&x&1\\ 0&0&0&0&-1&0&0&0\\ -1&0&0&0&0&0&0&0\\ 0&0&0&0&0&0&1&0\\ 0&0&0&-1&0&0&0&0\end{pmatrix}
$$
in the 24-dimensional space $H^0(Q_3,\text{C}(\varphi_4))$ has a zero-isomorphic cone. Hence $Q_3$ is generated by $Q_6$. Next, 
$$
\varphi_5=
\begin{pmatrix}0&0&0&1\\ -y^2&-x&0&0\\ 0&0&0&0\\ 0&0&0&0\end{pmatrix}
\oplus
\begin{pmatrix}0&0&0&-1\\ y^2&x&0&0\\ 0&0&0&0\\ 0&0&0&0\end{pmatrix}
$$
represents an element in the 2-dimensional space $H^0(Q_1,Q_7)$, and the element represented by
$$
\begin{pmatrix}0&-1&0&0&0&0&0&0\\ 0&0&0&-1&0&0&0&0\\ 0&0&-1&0&0&0&0&0\\ -x&0&0&0&0&0&0&0\\ 0&0&0&0&0&0&0&0\\ -1&0&0&0&0&0&0&0\\ 0&0&0&0&0&0&0&0\\ 0&0&0&0&0&0&0&0\end{pmatrix}
\oplus
\begin{pmatrix}0&1&0&0&0&0&0&0\\ 0&0&0&1&0&0&0&0\\ 0&0&1&0&0&0&0&0\\ x&0&0&0&0&0&0&0\\ 0&0&0&0&0&0&0&0\\ -1&0&0&0&0&0&0&0\\ 0&0&0&0&0&0&0&0\\ 0&0&0&0&0&0&0&0\end{pmatrix}
$$
in the 7-dimensional space $H^0(Q_4,\text{C}(\varphi_5))$ has a zero-isomorphic cone. Hence $Q_4$ is generated by $Q_1$ and $Q_7$. Finally, 
$$
\varphi_6=
\begin{pmatrix}x&0&y^2&0\\ 0&0&0&0\\ 0&0&0&1\\ 0&0&0&-y^2\end{pmatrix}
\oplus
\begin{pmatrix}-x&0&-y^2&0\\ 0&0&0&0\\ 0&0&0&-1\\ 0&0&0&y^2\end{pmatrix}
$$
represents an element in the 4-dimensional space $H^0(Q_4,Q_1)$, and the element represented by
$$
\begin{pmatrix}0&1&0&0&0&0&0&0\\ 0&0&0&1&0&0&0&0\\ 0&0&-1&0&0&0&0&0\\ y&0&0&0&x&0&0&0\\ 0&0&0&0&-1&0&0&0\\ 1&0&0&0&0&0&0&0\\ 0&0&0&0&0&0&0&0\\ 0&0&-y&0&0&-1&0&0\end{pmatrix}
\oplus
\begin{pmatrix}0&1&0&0&0&0&0&0\\ 0&0&0&1&0&0&0&0\\ 0&0&-1&0&0&0&0&0\\ y&0&0&0&x&0&0&0\\ 0&0&0&0&1&0&0&0\\ -1&0&0&0&0&0&0&0\\ 0&0&0&0&0&0&0&0\\ 0&0&y&0&0&1&0&0\end{pmatrix}
$$
in the 14-dimensional space $H^0(Q_5,\text{C}(\varphi_6))$ has a zero-isomorphic cone. Hence $Q_5$ is generated by $Q_1$ and $Q_4$. 

In summary, every D-brane in $\text{MF}(W_{\text{E}_7})$ can be viewed as a (repeated) tachyon condensation of the single brane $Q_7$: $\text{MF}(W_{\text{E}_7})=\text{tria}(Q_7)$.

\subsection*{ADE-type $\boldsymbol{\text{E}_8}$}

The category $\text{MF}(W_{\text{E}_8})$ with $W_{\text{E}_8}=x^3+y^5+z^2$ has a complete set of indecomposable objects $Q_i=(\begin{smallmatrix}0&f_i\\ f_i&0\end{smallmatrix})$ given by
\begin{align*}
f_1=&
\begin{pmatrix}
z & 0 & x & y \\
0 & z & y^{4} & -x^{2} \\
x^{2} & y & -z & 0 \\
y^{4} & -x & 0 & -z
\end{pmatrix} \, , \\
f_2=&
\begin{pmatrix}
z & -y^{2} & xy & 0 & -x^{2} & 0 \\
-y^{3} & -z & 0 & 0 & 0 & x \\
0 & 0 & -z & x & 0 & y \\
0 & -xy & x^{2} & z & y^{4} & 0 \\
-x & 0 & 0 & y & -z & 0 \\
0 & x^{2} & y^{4} & 0 & -xy^{3} & z
\end{pmatrix} \, , \\
f_3=&
\begin{pmatrix}
-z & 0 & -xy & y^{2} & 0 & 0 & x^{2} & 0 \\
0 & -z & y^{3} & 0 & 0 & 0 & 0 & x \\
0 & y^{2} & z & 0 & 0 & -x & 0 & 0 \\
y^{3} & xy & 0 & z & -x^{2} & 0 & 0 & 0 \\
0 & 0 & 0 & -x & -z & 0 & y^{3} & y \\
0 & 0 & -x^{2} & 0 & 0 & -z & 0 & y^{2} \\
x & 0 & 0 & 0 & y^{2} & -y & z & 0 \\
0 & x^{2} & 0 & 0 & 0 & y^{3} & 0 & z
\end{pmatrix} \, , \\
f_4=&
\begin{pmatrix}
z & 0 & xy & 0 & 0 & -y^{2} & y^{3} & 0 & -x^{2} & 0 \\
0 & -z & 0 & 0 & 0 & 0 & 0 & -y^{2} & 0 & x \\
0 & 0 & -z & y^{2} & 0 & 0 & 0 & x & 0 & 0 \\
0 & xy & y^{3} & z & 0 & 0 & -x^{2} & 0 & 0 & 0 \\
0 & y^{2} & 0 & 0 & z & -x & 0 & 0 & y^{3} & 0 \\
-y^{3} & 0 & 0 & 0 & -x^{2} & -z & 0 & 0 & 0 & y^{2} \\
0 & 0 & 0 & -x & 0 & 0 & -z & 0 & 0 & y \\
0 & -y^{3} & x^{2} & 0 & 0 & 0 & xy^{2} & z & 0 & 0 \\
-x & 0 & 0 & 0 & y^{2} & 0 & 0 & y & -z & 0 \\
0 & x^{2} & xy^{2} & 0 & 0 & 0 & y^{4} & 0 & 0 & z
\end{pmatrix} \, , \\
f_5=&
\begin{pmatrix}
-z & 0 & 0 & 0 & 0 & 0 & 0 & y^{2} & 0 & 0 & 0 & x \\
0 & -z & -xy & 0 & 0 & 0 & y^{3} & -y^{2} & 0 & 0 & x^{2} & 0 \\
0 & 0 & z & 0 & 0 & -y^{2} & 0 & 0 & y^{3} & -x & 0 & 0 \\
xy & 0 & 0 & z & -y^{3} & 0 & 0 & 0 & -x^{2} & 0 & 0 & 0 \\
0 & 0 & 0 & -y^{2} & -z & 0 & 0 & x & 0 & 0 & 0 & 0 \\
0 & 0 & -y^{3} & 0 & 0 & -z & -x^{2} & 0 & 0 & 0 & xy^{2} & y^{2} \\
y^{2} & y^{2} & 0 & 0 & 0 & -x & z & 0 & 0 & 0 & 0 & 0 \\
y^{3} & 0 & 0 & 0 & x^{2} & 0 & 0 & z & -xy^{2} & 0 & 0 & 0 \\
0 & 0 & 0 & -x & 0 & 0 & 0 & 0 & -z & 0 & 0 & y \\
0 & 0 & -x^{2} & -y^{3} & 0 & 0 & xy^{2} & 0 & 0 & -z & -y^{4} & 0 \\
0 & x & 0 & 0 & y^{2} & 0 & 0 & 0 & 0 & -y & z & 0 \\
x^{2} & 0 & 0 & 0 & -xy^{2} & 0 & 0 & 0 & y^{4} & 0 & 0 & z
\end{pmatrix} \, , \\
f_6=&
\begin{pmatrix}
-z & 0 & 0 & y^{2} & 0 & x \\
xy & z & -y^{3} & 0 & -x^{2} & 0 \\
0 & -y^{2} & -z & x & 0 & 0 \\
y^{3} & 0 & x^{2} & z & -xy^{2} & 0 \\
0 & -x & 0 & 0 & -z & y \\
x^{2} & 0 & -xy^{2} & 0 & y^{4} & z
\end{pmatrix} \, , \\
f_7=&
\begin{pmatrix}
z & 0 & 0 & 0 & -y^{3} & 0 & 0 & -x \\
xy & -z & 0 & 0 & 0 & y^{2} & x^{2} & 0 \\
0 & 0 & -z & y^{2} & 0 & x & -y^{3} & 0 \\
0 & 0 & 0 & z & -x^{2} & 0 & 0 & y^{2} \\
-y^{2} & 0 & 0 & -x & -z & 0 & 0 & 0 \\
0 & y^{3} & x^{2} & 0 & xy^{2} & z & 0 & 0 \\
0 & x & -y^{2} & 0 & 0 & 0 & z & y \\
-x^{2} & 0 & 0 & y^{3} & 0 & 0 & 0 & -z
\end{pmatrix} \, , \\
f_8=&
\begin{pmatrix}
z & 0 & x & y^{2} \\
0 & z & y^{3} & -x^{2} \\
x^{2} & y^{2} & -z & 0 \\
y^{3} & -x & 0 & -z
\end{pmatrix} \, .
\end{align*}
$\text{MF}(W_{\text{E}_8})$ is generated by $Q_8$. To see this, first note that 
$$
\varphi_1=
\begin{pmatrix}
0 &  0 & 0 &  1 \\ 
0 &  0 & -y & 0 \\ 
0 &  1 & 0 &  0 \\ 
-y & 0 & 0 &  0
\end{pmatrix}
\oplus
\begin{pmatrix}
0 & 0 &  0 & -1 \\ 
0 & 0 &  y & 0 \\ 
0 & -1 & 0 & 0 \\ 
y & 0 &  0 & 0
\end{pmatrix}
$$
represents an element in the 8-dimensional space $H^0(Q_8,Q_8)$, and the element represented by
$$
\begin{pmatrix}
0 &  0 &  0 &  -1 & 0 & 0 & 0 & 0 \\ 
0 &  0 &  y^2 & 0 &  0 & 0 & 0 & 0 \\ 
0 &  -1 & 0 &  0 &  0 & 0 & 0 & 0 \\ 
y^2 & 0 &  0 &  0 &  0 & 0 & 0 & 0 \\ 
0 &  0 &  0 &  0 &  0 & 0 & 0 & 0 \\ 
0 &  0 &  1 &  0 &  0 & 0 & 0 & 0 \\ 
0 &  0 &  0 &  0 &  0 & 0 & 0 & 0 \\ 
1 &  0 &  0 &  0 &  0 & 0 & 0 & 0
\end{pmatrix}
\oplus
\begin{pmatrix}
0 &  0 &  0 &  -1 & 0 & 0 & 0 & 0 \\ 
0 &  0 &  y^2 & 0 &  0 & 0 & 0 & 0 \\ 
0 &  -1 & 0 &  0 &  0 & 0 & 0 & 0 \\ 
y^2 & 0 &  0 &  0 &  0 & 0 & 0 & 0 \\ 
0 &  0 &  0 &  0 &  0 & 0 & 0 & 0 \\ 
0 &  0 &  -1 & 0 &  0 & 0 & 0 & 0 \\ 
0 &  0 &  0 &  0 &  0 & 0 & 0 & 0 \\ 
-1 & 0 &  0 &  0 &  0 & 0 & 0 & 0
\end{pmatrix}
$$
in the 4-dimensional space $H^0(Q_1, \text{C}(\varphi_1))$ has a zero-isomorphic cone. Hence $Q_1$ is generated by $Q_8$. Next, 
$$
\varphi_2=
\begin{pmatrix}
0 &   0 &  y & 0 \\ 
0 &   0 &  0 & xy \\ 
-xy & 0 &  0 & 0 \\ 
0 &   -y & 0 & 0
\end{pmatrix}
\oplus
\begin{pmatrix}
0 &  0 & -y & 0 \\ 
0 &  0 & 0 &  -xy \\ 
xy & 0 & 0 &  0 \\ 
0 &  y & 0 &  0
\end{pmatrix}
$$
represents another element in the 8-dimensional space $H^0(Q_8,Q_8)$, and the element represented by
$$
\begin{pmatrix}
0 & 0 &  -1 & 0 & 0 & 0 & 0 & 0 \\ 
0 & 0 &  0 &  0 & 0 & 0 & 0 & 1 \\ 
0 & 0 &  0 &  0 & 0 & 1 & 0 & 0 \\ 
0 & -1 & 0 &  0 & 0 & 0 & 0 & 0 \\ 
0 & 0 &  0 &  0 & 0 & 0 & 1 & 0 \\ 
0 & 0 &  0 &  1 & 0 & 0 & 0 & 0 \\ 
1 & 0 &  0 &  0 & 0 & 0 & 0 & 0 \\ 
0 & 0 &  0 &  0 & 1 & 0 & 0 & 0
\end{pmatrix}
\oplus
\begin{pmatrix}
0 & 0 & 1 & 0 & 0 & 0 &  0 & 0 \\ 
0 & 0 & 0 & 0 & 0 & 0 &  0 & -1 \\ 
0 & 0 & 0 & 0 & 0 & -1 & 0 & 0 \\ 
0 & 1 & 0 & 0 & 0 & 0 &  0 & 0 \\ 
0 & 0 & 0 & 0 & 0 & 0 &  1 & 0 \\ 
0 & 0 & 0 & 1 & 0 & 0 &  0 & 0 \\ 
1 & 0 & 0 & 0 & 0 & 0 &  0 & 0 \\ 
0 & 0 & 0 & 0 & 1 & 0 &  0 & 0
\end{pmatrix}
$$
in the 24-dimensional space $H^0(Q_3, \text{C}(\varphi_2))$ has a zero-isomorphic cone. Hence $Q_3$ is generated by $Q_8$. Next, 
$$
\varphi_3=
\begin{pmatrix}
0 &  -1 & 0 &  0 \\ 
xy & 0 &  0 &  0 \\ 
0 &  0 &  0 &  x \\ 
0 &  0 &  -y & 0
\end{pmatrix}
\oplus
\begin{pmatrix}
0 &  -1 & 0 &  0 \\ 
xy & 0 &  0 &  0 \\ 
0 &  0 &  0 &  x \\ 
0 &  0 &  -y & 0
\end{pmatrix}
$$
represents another element in the 8-dimensional space $H^0(Q_8,Q_8)$, and the element represented by
$$
\begin{pmatrix}
1 & 0 & 0 &  0 &  0 &  0 &  0 & 0 \\ 
0 & 0 & -x & 0 &  y^2 & 0 &  0 & 0 \\ 
0 & 0 & 0 &  0 &  0 &  -1 & 0 & 0 \\ 
0 & 0 & 0 &  -1 & 0 &  0 &  0 & 0 \\ 
0 & 0 & 0 &  0 &  0 &  0 &  0 & 0 \\ 
0 & 1 & 0 &  0 &  0 &  0 &  0 & 0 \\ 
0 & 0 & -1 & 0 &  0 &  0 &  0 & 0 \\ 
0 & 0 & 0 &  0 &  1 &  0 &  0 & 0
\end{pmatrix}
\oplus
\begin{pmatrix}
1 & 0 & 0 &  0 &  0 &  0 &  0 & 0 \\ 
0 & 0 & -x & 0 &  y^2 & 0 &  0 & 0 \\ 
0 & 0 & 0 &  0 &  0 &  -1 & 0 & 0 \\ 
0 & 0 & 0 &  -1 & 0 &  0 &  0 & 0 \\ 
0 & 0 & 0 &  0 &  0 &  0 &  0 & 0 \\ 
0 & 1 & 0 &  0 &  0 &  0 &  0 & 0 \\ 
0 & 0 & -1 & 0 &  0 &  0 &  0 & 0 \\ 
0 & 0 & 0 &  0 &  1 &  0 &  0 & 0
\end{pmatrix}
$$
in the 16-dimensional space $H^0(Q_6, \text{C}(\varphi_3))$ has a zero-isomorphic cone. Hence $Q_6$ is generated by $Q_8$. Next, 
$$
\varphi_4=
\begin{pmatrix}
0 &   -y & 0 &   0 \\ 
xy^2 & 0 &  0 &   0 \\ 
0 &   0 &  0 &   xy \\ 
0 &   0 &  -y^2 & 0
\end{pmatrix}
\oplus
\begin{pmatrix}
0 &   -y & 0 &   0 \\ 
xy^2 & 0 &  0 &   0 \\ 
0 &   0 &  0 &   xy \\ 
0 &   0 &  -y^2 & 0
\end{pmatrix}
$$
represents another element in the 8-dimensional space $H^0(Q_8,Q_8)$, and the element represented by
$$
\begin{pmatrix}
0 &  0 &  0 & 0 &  -1 & 0 &  0 &  0 \\ 
0 &  0 &  0 & 0 &  0 &  0 &  0 &  1 \\ 
0 &  0 &  0 & -1 & 0 &  0 &  0 &  0 \\ 
-1 & 0 &  0 & 0 &  0 &  0 &  0 &  0 \\ 
0 &  0 &  0 & 0 &  0 &  0 &  -1 & 0 \\ 
0 &  0 &  0 & 0 &  0 &  -1 & 0 &  0 \\ 
0 &  -1 & 0 & 0 &  0 &  0 &  0 &  0 \\ 
0 &  0 &  1 & 0 &  0 &  0 &  0 &  0
\end{pmatrix}
\oplus
\begin{pmatrix}
0 &  0 &  0 & 0 &  -1 & 0 &  0 &  0 \\ 
0 &  0 &  0 & 0 &  0 &  0 &  0 &  1 \\ 
0 &  0 &  0 & -1 & 0 &  0 &  0 &  0 \\ 
-1 & 0 &  0 & 0 &  0 &  0 &  0 &  0 \\ 
0 &  0 &  0 & 0 &  0 &  0 &  -1 & 0 \\ 
0 &  0 &  0 & 0 &  0 &  -1 & 0 &  0 \\ 
0 &  -1 & 0 & 0 &  0 &  0 &  0 &  0 \\ 
0 &  0 &  1 & 0 &  0 &  0 &  0 &  0
\end{pmatrix}
$$
in the 28-dimensional space $H^0(Q_7, \text{C}(\varphi_4))$ has a zero-isomorphic cone. Hence $Q_7$ is generated by $Q_8$. Next, 
$$
\varphi_5=
\begin{pmatrix}
0 &   -1 & 0 &   0 \\ 
xy^3 & 0 &  0 &   0 \\ 
0 &   0 &  0 &   x \\ 
0 &   0 &  -y^3 & 0
\end{pmatrix}
\oplus
\begin{pmatrix}
0 &   -1 & 0 &   0 \\ 
xy^3 & 0 &  0 &   0 \\ 
0 &   0 &  0 &   x \\ 
0 &   0 &  -y^3 & 0
\end{pmatrix}
$$
represents an element in the 4-dimensional space $H^0(Q_1,Q_1)$, and the element represented by
$$
\begin{pmatrix}
0 &  0 &  0 & 0 & -1 & 0 & 0 & 0 \\ 
0 &  -y & x & 0 & 0 &  0 & 0 & 0 \\ 
-1 & 0 &  0 & 0 & 0 &  0 & 0 & 0 \\ 
0 &  0 &  0 & 1 & 0 &  0 & 0 & 0 \\ 
0 &  0 &  0 & 0 & 0 &  0 & 0 & 0 \\ 
0 &  0 &  0 & 0 & 0 &  1 & 0 & 0 \\ 
0 &  0 &  1 & 0 & 0 &  0 & 0 & 0 \\ 
0 &  -1 & 0 & 0 & 0 &  0 & 0 & 0
\end{pmatrix}
\oplus
\begin{pmatrix}
0 &  0 &  0 & 0 & -1 & 0 & 0 & 0 \\ 
0 &  -y & x & 0 & 0 &  0 & 0 & 0 \\ 
-1 & 0 &  0 & 0 & 0 &  0 & 0 & 0 \\ 
0 &  0 &  0 & 1 & 0 &  0 & 0 & 0 \\ 
0 &  0 &  0 & 0 & 0 &  0 & 0 & 0 \\ 
0 &  0 &  0 & 0 & 0 &  1 & 0 & 0 \\ 
0 &  0 &  1 & 0 & 0 &  0 & 0 & 0 \\ 
0 &  -1 & 0 & 0 & 0 &  0 & 0 & 0
\end{pmatrix}
$$
in the 12-dimensional space $H^0(Q_2, \text{C}(\varphi_5))$ has a zero-isomorphic cone. Hence $Q_2$ is generated by $Q_1$. Next, 
$$
\varphi_6=
\begin{pmatrix}
0 &   0 &  0 &  x &   0 & 0 & 0 & 0 \\ 
0 &   0 &  0 &  -y^2 & 0 & 0 & 0 & 0 \\ 
-y^3 & 0 &  0 &  0 &   0 & 0 & 0 & 0 \\ 
0 &   0 &  0 &  0 &   0 & 0 & 0 & 0 \\ 
0 &   0 &  0 &  0 &   0 & 0 & 0 & 0 \\ 
0 &   0 &  y^3 & 0 &   0 & 0 & 0 & 0 \\ 
0 &   -1 & 0 &  0 &   0 & 0 & 0 & 0 \\ 
0 &   y^2 & 0 &  0 &   0 & 0 & 0 & 0
\end{pmatrix}
\oplus
\begin{pmatrix}
0 &   0 &  0 &  x &   0 & 0 & 0 & 0 \\ 
0 &   0 &  0 &  -y^2 & 0 & 0 & 0 & 0 \\ 
-y^3 & 0 &  0 &  0 &   0 & 0 & 0 & 0 \\ 
0 &   0 &  0 &  0 &   0 & 0 & 0 & 0 \\ 
0 &   0 &  0 &  0 &   0 & 0 & 0 & 0 \\ 
0 &   0 &  y^3 & 0 &   0 & 0 & 0 & 0 \\ 
0 &   -1 & 0 &  0 &   0 & 0 & 0 & 0 \\ 
0 &   y^2 & 0 &  0 &   0 & 0 & 0 & 0
\end{pmatrix}
$$
represents an element in the 8-dimensional space $H^0(Q_1,Q_3)$, and the element represented by
$$
{\begin{pmatrix}
0 &  0 & 0 & 0 & 0 & 0 &  0 &  0 & -1 & 0 & 0 & 0 & 0 & 0 & 0 & 0 \\ 
0 &  0 & x & 0 & 0 & -y & y^2 & 0 & 0 &  0 & 0 & 0 & 0 & 0 & 0 & 0 \\ 
-1 & 0 & 0 & 0 & 0 & 0 &  0 &  0 & 0 &  0 & 0 & 0 & 0 & 0 & 0 & 0 \\ 
0 &  0 & 0 & 0 & y & 0 &  0 &  1 & 0 &  0 & 0 & 0 & 0 & 0 & 0 & 0 \\ 
0 &  0 & 0 & 0 & 0 & 0 &  0 &  0 & 0 &  0 & 0 & 0 & 0 & 0 & 0 & 0 \\ 
0 &  0 & 0 & 0 & 0 & 0 &  0 &  0 & 0 &  0 & 0 & 0 & 0 & 0 & 0 & 0 \\ 
0 &  0 & 0 & 0 & 0 & 0 &  0 &  0 & 0 &  0 & 0 & 0 & 0 & 0 & 0 & 0 \\ 
0 &  0 & 0 & 0 & 0 & 0 &  0 &  0 & 0 &  0 & 0 & 0 & 0 & 0 & 0 & 0 \\ 
0 &  0 & 1 & 0 & 0 & 0 &  0 &  0 & 0 &  0 & 0 & 0 & 0 & 0 & 0 & 0 \\ 
0 &  1 & 0 & 0 & 0 & 0 &  0 &  0 & 0 &  0 & 0 & 0 & 0 & 0 & 0 & 0 \\ 
0 &  0 & 0 & 0 & 1 & 0 &  0 &  0 & 0 &  0 & 0 & 0 & 0 & 0 & 0 & 0 \\ 
0 &  0 & 0 & 1 & 0 & 0 &  0 &  0 & 0 &  0 & 0 & 0 & 0 & 0 & 0 & 0 \\ 
0 &  0 & 0 & 0 & 0 & 0 &  1 &  0 & 0 &  0 & 0 & 0 & 0 & 0 & 0 & 0 \\ 
0 &  0 & 0 & 0 & 0 & 1 &  0 &  0 & 0 &  0 & 0 & 0 & 0 & 0 & 0 & 0 \\ 
0 &  0 & 0 & 0 & 0 & 0 &  0 &  0 & 0 &  0 & 0 & 0 & 0 & 0 & 0 & 0 \\ 
0 &  0 & 0 & 0 & 0 & 0 &  0 &  0 & 0 &  1 & 0 & 0 & 0 & 0 & 0 & 0
\end{pmatrix}}^{\oplus 2}
$$
in the 40-dimensional space $H^0(Q_4, \text{C}(\varphi_6))$ has a zero-isomorphic cone. Hence $Q_4$ is generated by $Q_1$ and $Q_3$. Finally, 
$$
\varphi_7=
\begin{pmatrix}
0 & 0 &   xy & 0 &  0 &  0 \\
0 & 0 &   0 &  0 &  0 &  -x \\
0 & 0 &   0 &  0 &  xy & 0 \\
0 & -xy & 0 &  0 &  0 &  0 \\
x & 0 &   0 &  0 &  0 &  0 \\
0 & 0 &   0 &  xy & 0 &  0
\end{pmatrix}
\oplus
\begin{pmatrix}
0 & 0 &   xy & 0 &  0 &  0 \\
0 & 0 &   0 &  0 &  0 &  -x \\
0 & 0 &   0 &  0 &  xy & 0 \\
0 & -xy & 0 &  0 &  0 &  0 \\
x & 0 &   0 &  0 &  0 &  0 \\
0 & 0 &   0 &  xy & 0 &  0
\end{pmatrix}
$$
represents an element in the 16-dimensional space $H^0(Q_6,Q_6)$, and the element represented by $(\begin{smallmatrix}\Phi&0\\ 0&-\Phi\end{smallmatrix})$, where
$$
\Phi=
\begin{pmatrix}
1 & 0 & 0 & 0 & 0 &  0 & 0 & 0 & 0 &  0 & 0 & 0 \\
0 & 0 & 0 & 1 & 0 &  0 & 0 & 0 & 0 &  0 & 0 & 0 \\
0 & 0 & 0 & 0 & 1 &  0 & 0 & 0 & 0 &  0 & 0 & 0 \\
0 & 0 & 0 & 0 & 0 &  0 & 0 & 1 & 0 &  0 & 0 & 0 \\
0 & 0 & 0 & 0 & 0 &  0 & 0 & 0 & 1 &  0 & 0 & 0 \\
0 & 0 & 0 & 0 & 0 &  0 & 0 & 0 & 0 &  0 & 0 & 1 \\
0 & 0 & 1 & 0 & 0 &  0 & 0 & 0 & 0 &  0 & 0 & 0 \\
1 & 1 & 0 & 0 & 0 &  0 & 0 & 0 & 0 &  0 & 0 & 0 \\
0 & 0 & 0 & 0 & 0 &  0 & 1 & 0 & 0 &  0 & 0 & 0 \\
0 & 0 & 0 & 0 & 0 &  1 & 0 & 0 & -y & 0 & 0 & 0 \\
0 & 0 & 0 & 0 & 0 &  0 & 0 & 0 & 0 &  0 & 1 & 0 \\
0 & 0 & 0 & 0 & -y & 0 & 0 & 0 & 0 &  1 & 0 & 0
\end{pmatrix} \, ,
$$
in the 60-dimensional space $H^0(Q_5,\text{C}(\varphi_7))$ has a zero-isomorphic cone. Hence $Q_5$ is generated by $Q_6$. 

In summary, every D-brane in $\text{MF}(W_{\text{E}_8})$ can be viewed as a (repeated) tachyon condensation of the single brane $Q_8$: $\text{MF}(W_{\text{E}_8})=\text{tria}(Q_8)$.


\begin{thebibliography}{10}

\bibitem{a0703279}
P.~S. Aspinwall, \emph{Topological {D-B}ranes and {C}ommutative {A}lgebra},
  \href{http://www.arxiv.org/abs/hep-th/0703279}{[hep-th/0703279]}.

\bibitem{al0104}
P.~S. Aspinwall and A.~Lawrence, \emph{Derived {C}ategories and {Z}ero-{B}rane
  {S}tability}, JHEP \textbf{0108} (2001), 004,
  \href{http://www.arxiv.org/abs/hep-th/0104147}{[hep-th/0104147]}.

\bibitem{ag2001}
L.~L. Avramov and D.~R. Grayson, \emph{Resulutions and {C}ohomology over
  {C}omplete {I}ntersections}, Computations in {A}lgebraic {G}eometry with
  {M}acaulay 2 (D.~Eisenbud, ed.), Algorithms and Computations in Mathematics,
  vol.~8, Springer, 2001, pp.~131--178.

\bibitem{smithnormalform}
F.~Jr. Ayres, \emph{Smith {N}ormal {F}orm}, Schaum's {O}utline of {T}heory and
  {P}roblems of {M}atrices, Schaum, 1962, pp.~188--195.

\bibitem{bd0102018}
I.~Brunner and J.~Distler, \emph{Torsion {D}-{B}ranes in {N}ongeometrical
  {P}hases}, Adv. Theor. Math. Phys. \textbf{5} (2002), 265--309,
  \href{http://arxiv.org/abs/hep-th/0102018}{[hep-th/0102018]}.

\bibitem{bdm0106262}
I.~Brunner, J.~Distler, and R.~Mahajan, \emph{Return of the {T}orsion
  {D}-{B}ranes}, Adv. Theor. Math. Phys. \textbf{5} (2002), 311--352,
  \href{http://arxiv.org/abs/hep-th/0106262}{[hep-th/0106262]}.

\bibitem{bg0503}
I.~Brunner and M.~R. Gaberdiel, \emph{Matrix factorisations and permutation
  branes}, JHEP \textbf{0507} (2005), 012,
  \href{http://www.arxiv.org/abs/hep-th/0503207}{[hep-th/0503207]}.

\bibitem{bg0506}
I.~Brunner and M.~R. Gaberdiel, \emph{The matrix factorisations of the {D}-model}, J. Phys. A
  \textbf{38} (2005), 7901--7920,
  \href{http://www.arxiv.org/abs/hep-th/0506208}{[hep-th/0506208]}.

\bibitem{bhls0305}
I.~Brunner, M~Herbst, W.~Lerche, and B.~Scheuner, \emph{Landau-{G}inzburg
  {R}ealization of {O}pen {S}tring {TFT}}, JHEP \textbf{0611} (2003), 043,
  \href{http://www.arxiv.org/abs/hep-th/0305133}{[hep-th/0305133]}.

\bibitem{bhlw0408}
I.~Brunner, M.~Herbst, W.~Lerche, and J.~Walcher, \emph{Matrix {F}actorizations
  {A}nd {M}irror {S}ymmetry: {T}he {C}ubic {C}urve}, JHEP \textbf{0611} (2006),
  006, \href{http://www.arxiv.org/abs/hep-th/0408243}{[hep-th/0408243]}.

\bibitem{buchberger}
B.~Buchberger, \emph{Ein {A}lgorithmus zum {A}uffinden der {B}asiselemente des
  {R}est-klassenringes nach einem nulldimensionalen {P}olynomideal}, PhD thesis.

\bibitem{MFcohom}
N.~Carqueville, \emph{Computing {BRST}-cohomology algorithmically}, available
  at \href{http://nils.carqueville.net}{http://nils.carqueville.net}.

\bibitem{c0412149}
K.~J. Costello, \emph{Topological conformal field theories and {C}alabi-{Y}au
  categories}, \href{http://arxiv.org/abs/math.QA/0412149}{[math.QA/0412149]}.

\bibitem{d0011}
M.~R. Douglas, \emph{D-branes, categories and {$N=1$} supersymmetry}, J.~Math.
  Phys. \textbf{42} (2001), 2818--2843,
  \href{http://www.arxiv.org/abs/hep-th/0011017}{[hep-th/0011017]}.

\bibitem{eisenbudcommalg}
D.~Eisenbud, \emph{Commutative {A}lgebra: with a {V}iew {T}oward {A}lgebraic
  {G}eometry}, Graduate Texts in Mathematics, Springer, 1999.

\bibitem{err0508}
H.~Enger, A.~Recknagel, and D.~Roggenkamp, \emph{Permutation branes and linear
  matrix factorisations}, JHEP \textbf{0601} (2006), 087,
  \href{http://www.arxiv.org/abs/hep-th/0508053}{[hep-th/0508053]}.

\bibitem{fgk0609034}
S.~Fredenhagen, M.~R. Gaberdiel, and C.~A. Keller, \emph{Bulk induced boundary
  perturbations}, J.~Phys. A \textbf{40} (2007), F17,
  \href{http://arxiv.org/abs/hep-th/0609034}{[hep-th/0609034]}.

\bibitem{GM}
S.~I. Gelfand and Yu.~I. Manin, \emph{Methods of homological algebra}, second
  ed., Springer Monographs in Mathematics, Springer, 2002.

\bibitem{gjlw0512}
S.~Govindarajan, H.~Jockers, W.~Lerche, and N.~Warner, \emph{Tachyon
  {C}ondensation on the {E}lliptic {C}urve}, Nucl. Phys.~B \textbf{765} (2007),
  240--286, \href{http://arxiv.org/abs/hep-th/0512208}{[hep-th/0512208]}.

\bibitem{SingularBook}
G.-M. Greuel and G.~Pfister, \emph{A {S}ingular {I}ntroduction to {C}ommutative
  {A}lgebra}, Springer, 2002.

\bibitem{Singular}
G.-M. Greuel, G.~Pfister, and H.~Sch\"onemann, \emph{{\sc Singular} 3.0}, {A
  Computer Algebra System for Polynomial Computations}, Centre for Computer
  Algebra, University of Kaiserslautern, 2005,
  \href{http://www.singular.uni-kl.de}{http://www.singular.uni-kl.de}.

\bibitem{hhp}
M.~Herbst, K.~Hori, and D.~Page, to appear.

\bibitem{hll0405}
M.~Herbst, C.~I. Lazaroiu, and W.~Lerche, \emph{D-brane effective action and
  tachyon condensation in topological minimal models}, JHEP \textbf{0503}
  (2005), 078,
  \href{http://www.arxiv.org/abs/hep-th/0405138}{[hep-th/0405138]}.

\bibitem{hll0402}
M.~Herbst, C.~I. Lazaroiu, and W.~Lerche, \emph{Superpotentials, ${A}_{\infty}$ {R}elations and {WDVV}
  {E}quations for {O}pen {T}opological {S}trings}, JHEP \textbf{0502} (2005),
  071, \href{http://www.arxiv.org/abs/hep-th/0402110}{[hep-th/0402110]}.

\bibitem{h0401}
K.~Hori, \emph{Boundary {RG} {F}lows of {$N=2$} {M}inimal {M}odels}, Banff
  2003, {M}irror symmetry {V} (N.~Yui, S.-T. Yau, and J.D. Lewis, eds.), Am. Math. Soc., 2006,
  \href{http://www.arxiv.org/abs/hep-th/0401135}{[hep-th/0401135]},
  pp.~381--405.

\bibitem{kst0511}
H.~Kajiura, K.~Saito, and A.~Takahashi, \emph{Matrix {F}actorizations and
  {R}epresentations of {Q}uivers {II}: type {ADE} case}, Adv. in Math.
  \textbf{211} (2007), 327--362,
  \href{http://www.arxiv.org/abs/math.AG/0511155}{[math.AG/0511155]}.

\bibitem{kl0210}
A.~Kapustin and Y.~Li, \emph{D-branes in {L}andau-{G}inzburg {M}odels and
  {A}lgebraic {G}eometry}, JHEP \textbf{0312} (2003), 005,
  \href{http://www.arxiv.org/abs/hep-th/0210296}{[hep-th/0210296]}.

\bibitem{kl0306}
A.~Kapustin and Y.~Li, \emph{D-branes in {T}opological {M}inimal {M}odels: the
  {L}andau-{G}inzburg {A}pproach}, JHEP \textbf{0407} (2004), 045,
  \href{http://www.arxiv.org/abs/hep-th/0306001}{[hep-th/0306001]}.

\bibitem{keller1994}
B.~Keller, \emph{Deriving {DG} categories}, Ann. Scient. \'{E}c. Norm. Sup.
  \textbf{$\boldsymbol{4^e}$ s\'{e}rie} (1994), 63--102.

\bibitem{ks0610175}
C.~A. Keller and S.~Rossi, \emph{Boundary states, matrix factorisations and
  correlation functions for the {E}-models}, JHEP \textbf{0703} (2007), 038,
  \href{http://www.arxiv.org/abs/hep-th/0610175}{[hep-th/0610175]}.

\bibitem{lpp2002}
R.~Laza, G.~Pfister, and D.~Popescu, \emph{Maximal {C}ohen-{M}acaulay models
  over the cone of an elliptic curve}, J.~Algebra \textbf{253} (2002), 209.

\bibitem{l0610120}
C.~I. Lazaroiu, \emph{Generating the superpotential on a {D}-brane category:
  {I}}, \href{http://arxiv.org/abs/hep-th/0610120}{[hep-th/0610120]}.

\bibitem{l0102122}
C.~I. Lazaroiu, \emph{Generalized complexes and string field theory}, JHEP
  \textbf{0106} (2001), 052,
  \href{http://arxiv.org/abs/hep-th/0102122}{[hep-th/0102122]}.

\bibitem{l0102183}
C.~I. Lazaroiu, \emph{Unitarity, {D}-brane dynamics and {D}-brane categories}, JHEP
  \textbf{0112} (2001), 031,
  \href{http://arxiv.org/abs/hep-th/0102183}{[hep-th/0102183]}.

\bibitem{l0312}
C.~I. Lazaroiu, \emph{On the boundary coupling of topological {L}andau-{G}inzburg
  models}, JHEP \textbf{0505} (2005), 037,
  \href{http://www.arxiv.org/abs/hep-th/0312286}{[hep-th/0312286]}.

\bibitem{l0310337}
K.~Lef\`{e}vre-Hasegawa, \emph{Sur les {A}-infini cat\'egories}, PhD thesis,
  \href{http://arxiv.org/abs/math.CT/0310337}{[math.CT/0310337]}.

\bibitem{lvw1989}
W.~Lerche, C.~Vafa, and N.~Warner, \emph{Chiral {R}ings in $N=2$
  {S}uperconformal {T}heories}, Nucl. Phys.~B \textbf{324} (1989), 427.

\bibitem{m1989}
E.~J. Martinec, \emph{Algebraic {G}eometry and {E}ffective {L}agrangians},
  Phys. Lett.~B \textbf{217} (1989), 431.

\bibitem{Neeman}
A.~Neeman, \emph{Triangulated {C}ategories}, Annals of Mathematics Studies,
  Princeton University Press, 2001.

\bibitem{o0503}
D.~Orlov, \emph{Derived categories of coherent sheaves and triangulated
  categories of singularities},
  \href{http://arxiv.org/abs/math.AG/0503632}{[math.AG/0503632]}.

\bibitem{o0302}
D.~Orlov, \emph{Triangulated categories of singularities and {D}-branes in
  {L}andau-{G}inzburg models}, Tr. Mat. Inst. Steklova \textbf{246} (2004),
  240--262, \href{http://www.arxiv.org/abs/math.AG/0302304}{[math.AG/0302304]}.

\bibitem{qv}
A.~Quintero~V\'{e}lez, \emph{Mc{K}ay correspondence for {L}andau-{Ginzburg}
  models}, \href{http://arxiv.org/abs/0711.4774}{[arXiv:0711.4774]}.

\bibitem{rrs0003110}
A.~Recknagel, D.~Roggenkamp, and V.~Schomerus, \emph{Bulk induced boundary
  perturbations}, Nucl.~Phys. B \textbf{588} (2000), 552--564,
  \href{http://arxiv.org/abs/hep-th/0003110}{[hep-th/0003110]}.

\bibitem{s0511286}
M.~Schnabl, \emph{Analytic solution for tachyon condensation in open string
  field theory}, Adv. Theor. Math. Phys. \textbf{0511} (2006), 433--501,
  \href{http://www.arxiv.org/abs/hep-th/0511286}{[hep-th/0511286]}.

\bibitem{schreyer1987}
F.~O. Schreyer, \emph{Finite and countable {CM}-representation type},
  Singularities, Representations of Algebras, and Vector Bundles (G.-M. Greuel
  and G.~Trautmann, eds.), Lecture Notes in Mathematics, vol. 1273, Springer,
  1987, pp.~9--34.

\bibitem{s9805170}
A.~Sen, \emph{Tachyon {C}ondensation on the {B}rane {A}ntibrane {S}ystem}, JHEP
  \textbf{9808} (1998), 012,
  \href{http://www.arxiv.org/abs/hep-th/9805170}{[hep-th/9805170]}.

\bibitem{s9911116}
A.~Sen, \emph{Universality of the {T}achyon {P}otential}, JHEP \textbf{9912}
  (1999), 027,
  \href{http://www.arxiv.org/abs/hep-th/9911116}{[hep-th/9911116]}.

\bibitem{vw1989}
C.~Vafa and N.~Warner, \emph{Catastrophes and the classification of conformal
  theories}, Phys. Lett.~B \textbf{218} (1989), 51.

\bibitem{w9301}
E.~Witten, \emph{Phases of ${N}=2$ {T}heories in {T}wo {D}imensions}, Nucl.
  Phys.~B \textbf{403} (1993), 159--222,
  \href{http://www.arxiv.org/abs/hep-th/9301042}{[hep-th/9301042]}.

\bibitem{yoshino}
Y.~Yoshino, \emph{Cohen-{M}acaulay {M}odules over {C}ohen-{M}acaulay {R}ings},
  London Mathematical Society Lecture Note Series, Cambridge University Press,
  1990.

\end{thebibliography}
\end{document}